\documentclass[pra,twocolumn,amsmath,amssymb,superscriptaddress]{revtex4-1}

\usepackage{epsfig,amsmath}
\usepackage{subfigure}
\usepackage{graphicx}% Include figure files
\usepackage{dcolumn}% Align table columns on decimal point
\usepackage{stmaryrd}
\usepackage{mathrsfs}
\usepackage{pifont}
\usepackage{amsthm}
\usepackage{amssymb}
\usepackage{bm}
\usepackage{latexsym}
\usepackage[colorlinks=true,linkcolor=blue,citecolor=blue]{hyperref}
\usepackage{color}
\usepackage{epstopdf}

\begin{document}

\title{Suppressing classical noise in the accelerated geometric phase gate by optimized dynamical decoupling}

\author{Da-tong Chen}
\affiliation{School of Physics, Zhejiang University, Hangzhou 310027, Zhejiang, China}

\author{Jun Jing}
\email{jingjun@zju.edu.cn}
\affiliation{School of Physics, Zhejiang University, Hangzhou 310027, Zhejiang, China}

\begin{abstract}
In the quantum-computation scenario, the geometric phase-gates are becoming increasingly attractive for their intrinsic fault tolerance to disturbance. With an adiabatic cyclic evolution, Berry phase appears to realize a geometric transformation. Performing the quantum gates as many as possible within the timescale of coherence, however, remains an inconvenient bottleneck due to the systematic errors. Here we propose an accelerated adiabatic quantum gate based on the Berry phase, the transitionless driving, and the dynamical decoupling. It reconciles a high fidelity with a high speed in the presence of control noise or imperfection. We optimize the dynamical-decoupling sequence in the time domain under a popular Gaussian noise spectrum following the inversely quadratic power-law.
\end{abstract}

\maketitle

\section{Introduction}

By processing the data in a quantum-mechanical way, quantum computation out-performs its classical counterpart in selected algorithms or tasks~\cite{ShorAlgorithm,GroverAlgorithm}. The simulation of the coherent evolution of generic quantum systems can be modelled by performing ordered sequences of high-fidelity unitary operations~\cite{QuantumSimulation1,QuantumSimulation2,QuantumSimulation3}. The ubiquitous noises from uncontrollable environment and imprecise control, however, are inevitable in any experimental proposal, which pose challenges to the quality of quantum gate in terms of fidelity and operation time~\cite{QuantumNoise,QuantumNoise2}. The quantum geometric phase~\cite{BerryPhase,AAphase,GeometricPhase1,GeometricPhase2,GeometricPhase3}, which is accumulated as the system is moving along the parametric path, has intrinsic tolerant property against certain fluctuations during the trajectory. The geometric quantum computation is therefore appealing and becoming feasible in many experimental platforms, including the trapped ions~\cite{TrappedIon1,TrappedIon2}, the nuclear magnetic resonance~\cite{NMR1,NMR2}, and the superconducting circuits~\cite{SuperconductingCircuits1,SuperconductingCircuits2,RobustNGQC2,RobustNGQC1}.

As a benchmark implementation of the Abelian geometric transformation, the quantum gate based on the Berry phase requires that the quantum system undergoes a desired adiabatic loop in the parametric space~\cite{BerryPhase}. The process is intrinsically slow. Reducing the evolution time is in practice demanded to avoid the accumulation of the decoherence influence. On the other hand, however, speeding up always gives rise to the transition among the instantaneous eigenstates of system, which is supposed to be ``frozen'' in the adiabatic condition to maintain the gate fidelity. So that the competitive coherence time and the high fidelity are seemingly contradictory for the adiabatic geometric quantum gates. One of the compromise solutions to reduce the evolution time is using the quantum gates based on the Aharonov-Anandan phase~\cite{AAphase,AAphaseQuantumGate}, which emerges in arbitrary cyclic unitary evolutions. Alternatively, one can accelerate the evolution process via the transitionless quantum driving (TQD)~\cite{TQD}, which is formulated by the ancillary Hamiltonian to avoid the level crossing during the time evolution.

The TQD-accelerated quantum gate based on the Berry phase is still sensitive to the systematic errors in the control Hamiltonian or driving parameters. A significant feature of geometric quantum computation is its compatibility with certain error-suppression techniques, e.g., the dynamical decoupling (DD)~\cite{DD1,MultipleCPMG,UDD1,UDD2,GaussianNoiseForm2}, the unconventional geometric gates~\cite{UnconventionalGQC1,UnconventionalGQC2,UnconventionalGQC3}, the decoherence-free subspaces~\cite{DecoherenceFreeSubspace}, the quantum error correction~\cite{QuantumErrorCorrection1,QuantumErrorCorrection2}, and the dynamical correction~\cite{RobustNGQC1,RobustNGQC2}, to maintain the fidelity of the adiabatic or nonadiabatic unitary transformation. The dynamical decoupling was originated in the high-precision magnetic resonance spectroscopy and has been applied to quantum control as a long standing technique. A typical DD is to neutralize the effects from the fluctuation noises by applying a sequence of inverse operations to a two-level system~\cite{DD1}. In this work, we aim to find an optimized dynamical-decoupling sequence to protect the TQD-accelerated Berry-phase gate.

Particularly, we consider the effect from the Gaussian stochastic noises~\cite{StochasticNoise} in the control parameters of our nonadiabatic geometric quantum gates, which would deviate the output states from the noise-free result. Under various resource of noises, we analyze the robustness of the gate-fidelity for a general input state to estimate the deviation and then apply the dynamical-decoupling sequences to improve the fidelity. In the time domain, we analytically derive an optimized DD sequence under the Gaussian color noise following the inversely quadratic power-law spectrum.

The rest of the work is arranged as follows. In Sec.~\ref{secModel}, we start from a semi-classical Rabi model under parametric driving and obtain a general time-dependent qubit Hamiltonian corrected by the counter-rotating interaction up to the first order. This Hamiltonian is used to establish a set of more accurate Berry phase gates accelerated by the transitionless quantum driving. In Sec.~\ref{secNoise}, we consider the stochastic fluctuations in various control parameters and their damages to the fidelity of the preceding nonadiabatic transformation in the free induction decay (FID). Then in Sec.~\ref{secDD}, we apply the DD technique to the Berry phase gate and discuss the results under the inversely quadratic noise spectrum. The detailed derivation of the optimized control sequence can be found in Appendix~\ref{appendix}. We discuss and summarize the whole work in Sec.~\ref{secConclusion}.

\section{A more accurate effective Hamiltonian}\label{secModel}

We start with a Rabi model under parametric driving, in which a two-level system (qubit) with energy-spacing $\omega_a$ is controlled by a driving field with time-modulated frequency $\omega_b(t)$ and phase factor $\varphi_R(t)$. The strength of the dipole-dipole interaction between the qubit and the driving field (Rabi-frequency) is described by $\Omega_R(t)$. The system Hamiltonian thus can be written as
\begin{equation}\label{Hinitial}
H(t)=\frac{\omega_a}{2}\sigma_z+\Omega_R(t)\cos\left[\Omega_b(t)+\varphi_R(t)\right]\sigma_x,
\end{equation}
where $\Omega_b(t)\equiv\int_0^tds\,\omega_b(s)$. It is not convenient to construct a credible quantum gate directly from the original Hamiltonian in Eq.~(\ref{Hinitial}) for the lack of a compact expression of the time evolution operator. It is popular to see the application of the rotating-wave approximation (RWA) in the previous treatments~\cite{RabiModel,RWA,TrappedIon2}. With respect to the unitary transformation $R(t)=\exp[i\Omega_b(t)\sigma_z/2]$, one can find
\begin{equation}\label{HRWA}
\begin{aligned}
&H'(t)=R(t)H(t)R^\dagger(t)+i\dot{R}(t)R^\dagger(t)\\
&=\frac{1}{2}\big\{[\omega_a-\omega_b(t)]\sigma_z\\
&+\Omega_R(t)[\cos\varphi_R(t)+\cos(2\Omega_b(t)+\varphi_R(t))]\sigma_x\\
&+\Omega_R(t)[\sin\varphi_R(t)-\sin(2\Omega_b(t)+\varphi_R(t))]\sigma_y\big\}\\
&\approx\frac{\omega_a-\omega_b(t)}{2}\sigma_z+\frac{\Omega_R(t)}{2}\left[e^{-i\varphi_R(t)}\sigma_++h.c.\right],
\end{aligned}
\end{equation}
where the terms with the high-frequency $2\Omega_b(t)$ are omitted. The error between the resultant Hamiltonian under RWA and the original one is thus in the first order of $\mathcal{O}(\Omega_R)$. $H'(t)$ applies to the dispersive regime of a sufficiently weak driving strength. The contribution of the omitted counter-rotating terms becomes however significant in the strong-coupling regime and demonstrates intriguing dynamical behaviors~\cite{StrongDrivingDynamics1,StrongDrivingDynamics2,StrongDrivingDynamics3,StrongDrivingDynamics4}.

To retain the first-order contribution from the counter-rotating terms, we choose a different approach with a modified rotating-wave approximation~\cite{StrongDrivingDynamics3}. We come back to the Hamiltonian in Eq.~(\ref{Hinitial}) and apply a unitary transformation with respect to
\begin{equation}\label{Soft}
S(t)=\exp\left\{i\frac{\Omega_R(t)}{\omega_a+\omega_b(t)}\sin[\Omega_b(t) +\varphi_R(t)]\sigma_x\right\}.
\end{equation}
The original Hamiltonian in the interaction picture is then rewritten as
\begin{equation}\label{HS}
\begin{aligned}
&H_0(t)=S(t)H(t)S^\dagger(t)+i\dot{S}(t)S^\dagger(t)\\
&=\frac{\omega_a}{2}\sigma_z+\Omega_R(t)\cos\left[\Omega_b(t)+\varphi_R(t)\right]\sigma_x \\
&+\frac{\Omega_R(t)\omega_a}{\omega_a+\omega_b(t)}\sin\left[\Omega_b(t)+\varphi_R(t)\right]\sigma_y+\mathcal{O}(\Omega_R^2) \\&-\frac{d}{dt}\left\{\frac{\Omega_R(t)}{\omega_a+\omega_b(t)}\sin[\Omega_b(t)+\varphi_R(t)]\sigma_x\right\} \\
&\approx\frac{\omega_a}{2}\sigma_z+\frac{\Omega_R(t)\omega_a}{\omega_a+\omega_b(t)}\left[e^{-i\Omega_b(t)
-i\varphi_R(t)}\sigma_++h.c.\right],
\end{aligned}
\end{equation}
where we omitted the contribution up to the second-order of $\mathcal{O}(\Omega_R^2)$ and the first-order derivative of the driving parameters with respect to time~\cite{RWA,TrappedIon2} under the assumptions that $|\dot{\omega}_b(t)|$, $|\dot{\varphi}_R(t)|$, $|\dot{\Omega}_R(t)|\ll|\omega_a+\omega_b(t)|$.

\begin{figure}
\includegraphics[width=0.4\textwidth]{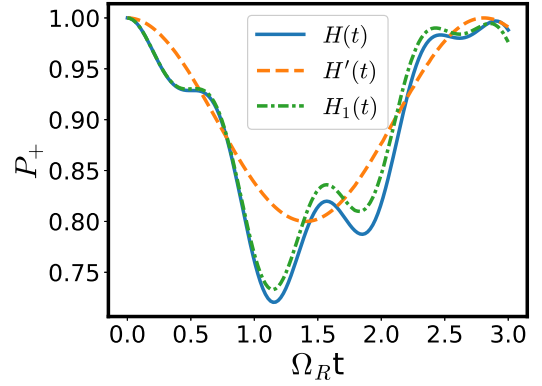}
\caption{The population dynamics of the qubit on $|+\rangle$ under various Hamiltonians, in units of the Rabi frequency $\Omega_R$. The parameters are set as $\omega_a=100\pi$ MHz, $\omega_b=60\pi$ MHz, and $\Omega_R=20\pi$ MHz.}
\label{FigCompare}
\end{figure}

Subsequently, in the rotating frame with respect to $R(t)$, one can find a standard time-modulated Hamiltonian describing a qubit under an effective 3D magnetic field, i.e.,
\begin{equation}\label{Htransformed}
\begin{aligned}
&H_1(t)=R(t)H_0(t)R^\dagger(t)+i\dot{R}(t)R^\dagger(t)\\
&=\frac{\vec{B}(t)}{2}\cdot\vec{\sigma}=\frac{B_0(t)}{2}\vec{n}(t)\cdot\vec{\sigma},
\end{aligned}
\end{equation}
where $\vec{n}(t)\equiv[\sin\theta(t)\cos\phi(t), \sin\theta(t)\sin\phi(t), \cos\theta(t)]$ parameterizes the direction of the magnetic field and $\vec{\sigma}$ represents the vector of Pauli operators. Using the driving parameters, we have
\begin{equation}\label{Bphitheta}
\left\{
\begin{aligned}
B_0(t)&=\sqrt{\left[\omega_a-\omega_b(t)\right]^2+\left[\frac{2\Omega_R(t)\omega_a}{\omega_a+\omega_b(t)}\right]^2},\\
\phi(t)&=\varphi_R(t),\\
\theta(t)&=\arctan\left[\frac{2\Omega_R(t)\omega_a}{\omega_a^2-\omega_b^2(t)}\right].
\end{aligned}
\right.
\end{equation}

The Hamiltonian in Eq.~(\ref{Htransformed}) adapts to a larger Rabi frequency $\Omega_R$ than that in Eq.~(\ref{HRWA}) by holding the first-order contribution from the counter-rotating interaction. The distinction between these two approximated Hamiltonians can be transparently illustrated by the Rabi oscillation of the population on the state $|+\rangle\equiv(1,0)^T$ for a two-level system, when the magnitudes of $\omega_a$, $\omega_b$, and $\Omega_R$ are chosen in almost the same order. The blue solid line, the yellow dashed line, and the green dot-dashed line in Fig.~\ref{FigCompare} represent the respective results under the original Hamiltonian $H(t)$ in Eq.~(\ref{Hinitial}), the RWA Hamiltonian $H'(t)$ in Eq.~(\ref{HRWA}), and the modified-RWA Hamiltonian $H_1(t)$. Much to our anticipation, the rotating-wave interaction captures the sinusoid behavior in quality, while losing nearly all the details of the dynamics. In contrast, with the aid of the unitary transformation $S(t)$ in Eq.~(\ref{Soft}), the analytical result by $H_1(t)$ is highly close to the numerical one by the original Hamiltonian $H(t)$.

We then apply the effective Hamiltonian $H_1(t)$ to build a more accurate quantum gate than $H'(t)$. Universal single-qubit gates can be implemented by virtue of the time-dependent $\theta(t)$ and $\phi(t)$ in Eq.~(\ref{Bphitheta}). $B_0$ is set as a constant number for simplicity. To avoid the undesired transition among the instantaneous eigenstates of $H_1(t)$ with accelerated time-dependence, one can add a counterdiabatic term following the transitionless quantum driving~\cite{TQD} approach. The ancillary Hamiltonian $H_{\rm CD}(t)$ can be expressed by
\begin{equation}\label{HTQD}
\begin{aligned}
&H_{\rm CD}(t)= \frac{1}{2}\left[\vec{n}(t)\times\dot{\vec{n}}(t)\right]\cdot\vec{\sigma}\\
&= \frac{1}{2}\big(-\dot{\theta} \sin \phi-\dot{\phi}\sin\theta\cos\theta\cos\phi,\\
& \dot{\theta}\cos\phi-\dot{\phi}\sin\theta\cos\theta\sin\phi, \dot{\phi}\sin^2\theta\big)\cdot\vec{\sigma}, \\
\end{aligned}
\end{equation}
where $\vec{n}(t)$ is the unit vector of the magnetic field in Eq.~(\ref{Htransformed}). In Eq.~(\ref{HTQD}), the explicit $t$-dependence of quantities has been omitted for simplicity, i.e., $\theta\equiv\theta(t)$ and $\phi\equiv\phi(t)$. The corrected Hamiltonian with the counterdiabatic term is $H_{\rm tot}(t)=H_1(t)+H_{\rm CD}(t)$, which could be diagonalized with the unitary transformation $U_0(t)=\exp(i\theta\sigma_y/2)\exp(i\phi\sigma_z/2)$:
\begin{equation}\label{HZ}
\begin{aligned}
& H_z(t)=U_0(t)H_{\rm tot}(t) U_0^\dagger(t)+i\dot{U}_0(t)U_0^\dagger(t)\\
&= \frac{1}{2}\left(B_0-\dot{\phi}\cos\theta\right)\sigma_z.
\end{aligned}
\end{equation}
Rotating back to the concatenated rotating frame with respect to $S(t)$ and $R(t)$, where $H_1(t)$ and $H_{\rm tot}(t)$ live, the time-evolution operator reads,
\begin{equation}
\begin{aligned}
&U(t)=U_0^\dagger(t)U_z(t)=e^{-i\phi\sigma_z/2}e^{-i\theta\sigma_y/2}e^{-i\int_0^tdsH_z(s)}\\
&=\begin{pmatrix}
\cos\frac{\theta(t)}{2}e^{-\frac{i}{2}[\alpha(t)+\phi(t)]} & -\sin\frac{\theta(t)}{2}e^{\frac{i}{2}[\alpha(t)-\phi(t)]}\\
\sin\frac{\theta(t)}{2}e^{-\frac{i}{2}[\alpha(t)-\phi(t)]} & \cos\frac{\theta(t)}{2}e^{\frac{i}{2}[\alpha(t)+\phi(t)]}
\end{pmatrix}
\end{aligned}
\end{equation}
with $U_z(t)\equiv\exp[-i\int_0^tdsH_z(s)]$ and $\alpha(t)=\int_0^tds(B_0-\dot{\phi}\cos\theta)$. Under a proper boundary condition, $U(t)$ can be used to realize any desired rotation or qubit-gate. For example, under the setting that $\theta(T)=\pi$ and $\alpha(T)-\phi(T)=\pi$, $U(T)$ realizes a Pauli-X gate up to a global phase. In the following discussion, we are concerned with the phase shift in a cyclic evolution. With $\theta(T)=2\pi$, the final time-evolution operator is found to be in a diagonal form:
\begin{equation}\label{idealUT}
U(T)=\begin{pmatrix}
        e^{-\frac{i}{2}[\alpha(T)+\phi(T)]} & 0\\
        0& e^{\frac{i}{2}[\alpha(T)+\phi(T)]}
    \end{pmatrix},
\end{equation}
up to an unobservable global $\pi$-phase. In particular, when $|\psi(0)\rangle=|\pm\rangle$ [$|-\rangle\equiv(0,1)^T$], one can derive a cyclic process $|\psi(T)\rangle =e^{i\pi\mp\frac{i}{2}[\alpha(T)+\phi(T)]}|\psi(0)\rangle= e^{i(\pi+\gamma_\pm)}|\psi(0)\rangle$. The total phase accumulated in this period is
\begin{equation}\label{Totalphase}
\gamma_\pm=\mp\frac{1}{2}\int_0^Tdt\left[B_0-\dot{\phi}(\cos\theta-1)\right].
\end{equation}
The dynamical phase could be obtained by the corrected Hamiltonian $H_{\rm tot}(t)$ and the eigenstates of the effective Hamiltonian $H_1(t)$. We have
\begin{equation}\label{dynamicalphase}
\begin{aligned}
& \gamma_{\pm}^d\equiv-\int_0^Tdt\langle\psi(t)|H_{\rm tot}(t)|\psi(t)\rangle \\
&=-\int_0^Tdt\langle\pm|U_0(t)H_{\rm tot}(t)U_0^{\dagger}(t)|\pm\rangle \\
&=-\int_0^Tdt\langle\pm|\frac{1}{2}\left(-\dot{\phi}\sin\theta\sigma_x+\dot{\theta}\sigma_y+B_0\sigma_z\right)|\pm\rangle \\ &=\mp\frac{1}{2}\int_0^TdtB_0=\mp\frac{1}{2}B_0T.
\end{aligned}
\end{equation}
The geometric phase is thus given by
\begin{equation}\label{AAphase}
\gamma_{\pm}^g=\gamma_{\pm}-\gamma_{\pm}^d=\mp\frac{1}{2}\int_0^Tdt\dot{\phi}(1-\cos\theta),
\end{equation}
which is exactly the solid angle in the Bloch sphere described by $\theta$ and $\phi$ up to a scale of $-1/2$~\cite{Solidangle}. The evolution period $T$ could be shortened by TQD as long as the boundary conditions are satisfied.

Provided that the dynamical phase is completely cancelled by any control technique, such as the dynamical decoupling~\cite{AAphaseQuantumGate,AAphaseQuantumGate2} to be discussed later, the geometric phase in Eq.~(\ref{AAphase}) determines the final time evolution operator:
\begin{equation}\label{QuantumGate}
U(T)\simeq
\begin{pmatrix}
e^{i\gamma_+^g } & 0\\
0& e^{i\gamma_-^g}
\end{pmatrix}
\simeq
\begin{pmatrix}
1 & 0\\
0& e^{i\int_0^T dt\dot{\phi}(1-\cos\theta)}
\end{pmatrix}.
\end{equation}
Then we can build a quantum geometric phase gate. Under the assumption $\theta(t)=\omega t$, the period is $T=2\pi/\omega$ and $U(t)={\rm diag}\{1, e^{i\eta}\}$, where $\eta=\gamma_-^g-\gamma_+^g$. Arbitrary phase gate could be attained by adjusting $\phi$. For example, we can have a Pauli-$Z$ gate when $\phi=\omega t/2$ and we have a $\pi/2$ gate when $\phi=\omega t/4$. In general, an arbitrary input state $|\psi(0)\rangle=\cos\alpha_1|+\rangle+\sin\alpha_1e^{i\alpha_2}|-\rangle$ would be transformed by the phase-gate to
\begin{equation}\label{FinalState}
|\psi(T)\rangle=\cos\alpha_1|+\rangle + \sin\alpha_1e^{i\alpha_2}e^{i\eta}|-\rangle
\end{equation}
where $\alpha_1$ and $\alpha_2$ are real-number angles with $\{\alpha_1, \alpha_2\}\in [0,2\pi]$.

\section{The effects from classical noise}\label{secNoise}

The geometric evolution in Eq.~(\ref{FinalState}), which has been accelerated by the counterdiabatic term, cannot be faithful in the presence of nonideal driving. The existence of the classical noise gives rise to the deviation of the gate-fidelity. Here we demonstrate how the noises in various control parameters would be detrimental to the performance of quantum gates in the absence of dynamical decoupling. Then in this section, we temporarily include the dynamical phase in Eq.~(\ref{dynamicalphase}) that is exclusively determined by $B_0$. The phase $\eta=\gamma_-^g-\gamma_+^g$ in Eq.~(\ref{FinalState}) then becomes $\eta=\gamma_--\gamma_+$, i.e.,
\begin{equation}\label{fullphase}
\eta=\int_0^Tdt\left[B_0-\dot{\phi}(\cos\theta-1)\right].
\end{equation}
To measure the phase deviation induced by the classical noise on the quantum gate, we use a gate-fidelity~\cite{PulseSequenceAnalysis2,GaussianNoiseForm2,PulseSequenceAnalysis1} defined as
\begin{equation}\label{Fidelity}
\mathcal{F}=\frac{1}{4\pi^2}\int_0^{2\pi}d\alpha_1\int_0^{2\pi}d\alpha_2
\frac{M\left[\langle+|\rho(T)|-\rangle\right]}{\langle+|\psi(T)\rangle\langle\psi(T)|-\rangle},
\end{equation}
where $\rho(t)$ is the density matrix under the time evolution operator $U(t)$ with noisy parameters and the wavefunction $|\psi(T)\rangle$ describes the ideal evolution in Eq.~(\ref{FinalState}). $M[\cdot]$ means the ensemble average over the random realizations of fluctuated control parameters, e.g., $B_0\rightarrow B_0+\delta_B(t)$ and the arbitrary input states featured with $\alpha_1$ and $\alpha_2$.

In this work, the stationary Gaussian noise~\cite{GaussianNoiseForm1,GaussianNoiseForm2} $\delta_{\xi}(t)$ is assumed to follow the statistical properties
\begin{equation}\label{Statistic}
\langle\delta_{\xi}(t)\rangle = 0,\quad
C(t-s)=\langle\delta_{\xi}(t)\delta_{\xi}(s)\rangle=\frac{\Gamma\gamma}{2}e^{-\gamma|t-s|},
\end{equation}
where $\xi$ indicates the noise source, $\Gamma$ is the correlation intensity of the noise, and $\gamma$ is the memory parameter. The Fourier transform of the two-point correlation function $C(t-s)$ gives rise to an inverse-quadratic spectral density,
\begin{equation}
S(\omega)=\int_{-\infty}^\infty dte^{i\omega t}C(t)=\frac{2\Gamma\gamma^2}{\omega^2+\gamma^2}.
\end{equation}
When $\gamma\rightarrow\infty$, $S(\omega)$ becomes structureless and describes a Markovian or white noise; while when $\gamma\rightarrow0$, it describes a typical non-Markovian noise with a finite-memory capability.

In the rest part of this section, we calculate the gate-fidelity during the free induction decay under the parametric fluctuations associated respectively with the magnetic field intensity $B_0$ and the phase derivative $\dot{\phi}$. Note these two parameters are separable in the expression of quantum phases, allowing the individual addressing over different noise resource. In comparison to both $B_0$ and $\dot{\phi}$, $\theta$ comes into the phase in a cosine function, see e.g., Eq.~(\ref{fullphase}), yielding a higher-order contribution. Thus the effect from noisy $\theta$ could be omitted. 

\subsection{Fidelity under noisy magnetic field intensity}\label{secNoiseB}

\begin{figure}[htbp]
\includegraphics[width=0.4\textwidth]{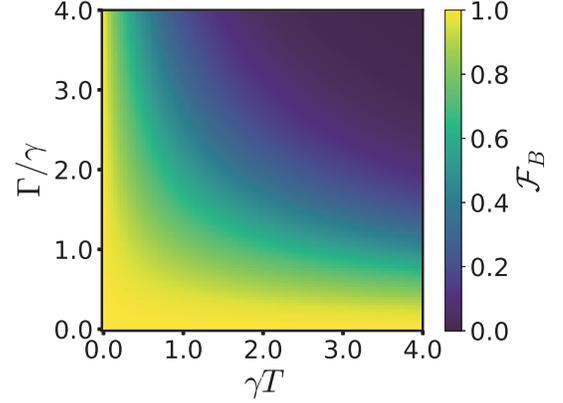}
\caption{Landscape of the gate-fidelity $\mathcal{F}_B$ under the stochastic $B_0$ in the parameter space of the strength-memory ratio $\Gamma/\gamma$ and the running period $\gamma T$.} \label{FigFidelityB}
\end{figure}

We first consider a fluctuated magnetic field $B_0$ in constructing the phase gate, which is determined by the driving parameters in Eq.~(\ref{Htransformed}). With respect to the unitary transformation $U_0(t)$, the nonideal counterdiabatic corrected Hamiltonian in Eq.~(\ref{HZ}) becomes
\begin{equation}\label{HnoiseB}
\tilde{H}_z(t)=\frac{1}{2}\left[B_0+\delta_B(t)-\dot{\phi}\cos\theta\right]\sigma_z.
\end{equation}
Consequently, the off-diagonal term of the final density matrix in Eq.~(\ref{Fidelity}) can be obtained by
\begin{equation}\label{plusrhoTminus}
\langle+|\rho(T)|-\rangle=\left\langle+\left|\tilde{U}^\dagger(T)\rho(0) \tilde{U}(T)\right|-\right\rangle
\end{equation}
where $\rho(0)\equiv|\psi(0)\rangle\langle\psi(0)|$ and $\tilde{U}(T)$ is
\begin{equation}
\tilde{U}(T)=e^{-i\phi(T)\sigma_z/2}e^{-i\theta(T)\sigma_y/2}e^{-i\int_0^Tds\tilde{H}_z(s)}.
\end{equation}
Then we have
\begin{equation}
\frac{\langle+|\rho(T)|-\rangle}{\langle+|\psi(T)\rangle\langle\psi(T)|-\rangle}=\exp\left[-i\int_0^Tdt\delta_B(t)\right].
\end{equation}
Note this result is independent of $\alpha_1$ and $\alpha_2$, meaning the noise effect on the gate-fidelity does not rely on the input states. Substituting it into Eq.~(\ref{Fidelity}) and using the statistical properties in Eq.~(\ref{Statistic}), we have~\cite{StochasticNoise}
\begin{equation}\label{FidelityB}
\mathcal{F}_{B}=e^{-\frac{1}{2}\int_0^Tdt\int_0^TdsC(t-s)}=
e^{-\frac{\Gamma}{2\gamma}\left(\gamma T+e^{-\gamma T}-1\right)}.
\end{equation}
The dependence of the gate-fidelity $\mathcal{F}_{B}$ on the dimensionless cyclic period $\gamma T$ and the memory parameter $\Gamma/\gamma$ is plotted in Fig.~\ref{FigFidelityB}.

From both Eq.~(\ref{FidelityB}) and Fig.~\ref{FigFidelityB}, a sufficiently large $\Gamma/\gamma$ or a sufficiently small $\gamma$ gives rise to a nearly exponential decay, which is consistent with a typical Markovian dynamics describing decoherence induced by the white noise. While the non-exponential decay appears in the short-time regime with $\gamma T\ll1$, where up to the leading order the exponent of the fidelity in Eq.~(\ref{FidelityB}) becomes quadratic to the running time:
\begin{equation}\label{ApproximatedFIdelityB}
\mathcal{F}_B\approx\exp\left(-\frac{\Gamma\gamma}{4}T^2\right).
\end{equation}
Then the characteristic decoherence time $T_2$ can be obtained by the definition $\ln \mathcal{F}_{B}(T_2)=-1$~\cite{PulseSequenceAnalysis2}
\begin{equation}\label{T2B}
T_2=\frac{2}{(\Gamma\gamma)^{1/2}}
\end{equation}
for the free induction decay. Clearly a high-level gate-fidelity can be maintained in the presence of a weak and non-Markovian classical noise. It is found when $\Gamma/\gamma\leq1.0$, $\mathcal{F}_B$ is over $0.90$ at $\gamma T=0.72$.

\subsection{Fidelity under noisy control phase $\phi$}\label{secNoisephi}

\begin{figure}[htbp]
\includegraphics[width=0.4\textwidth]{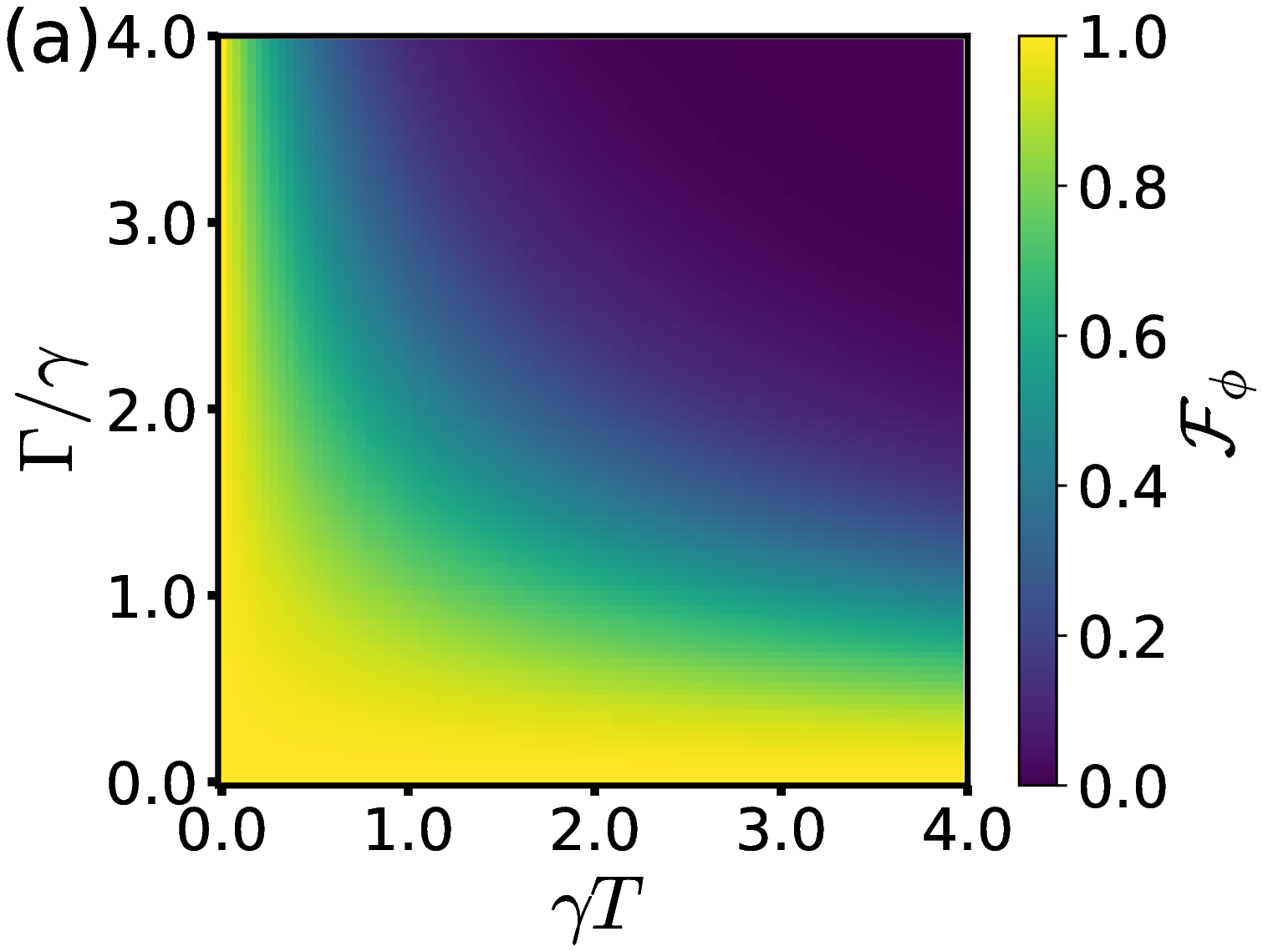}
\includegraphics[width=0.4\textwidth]{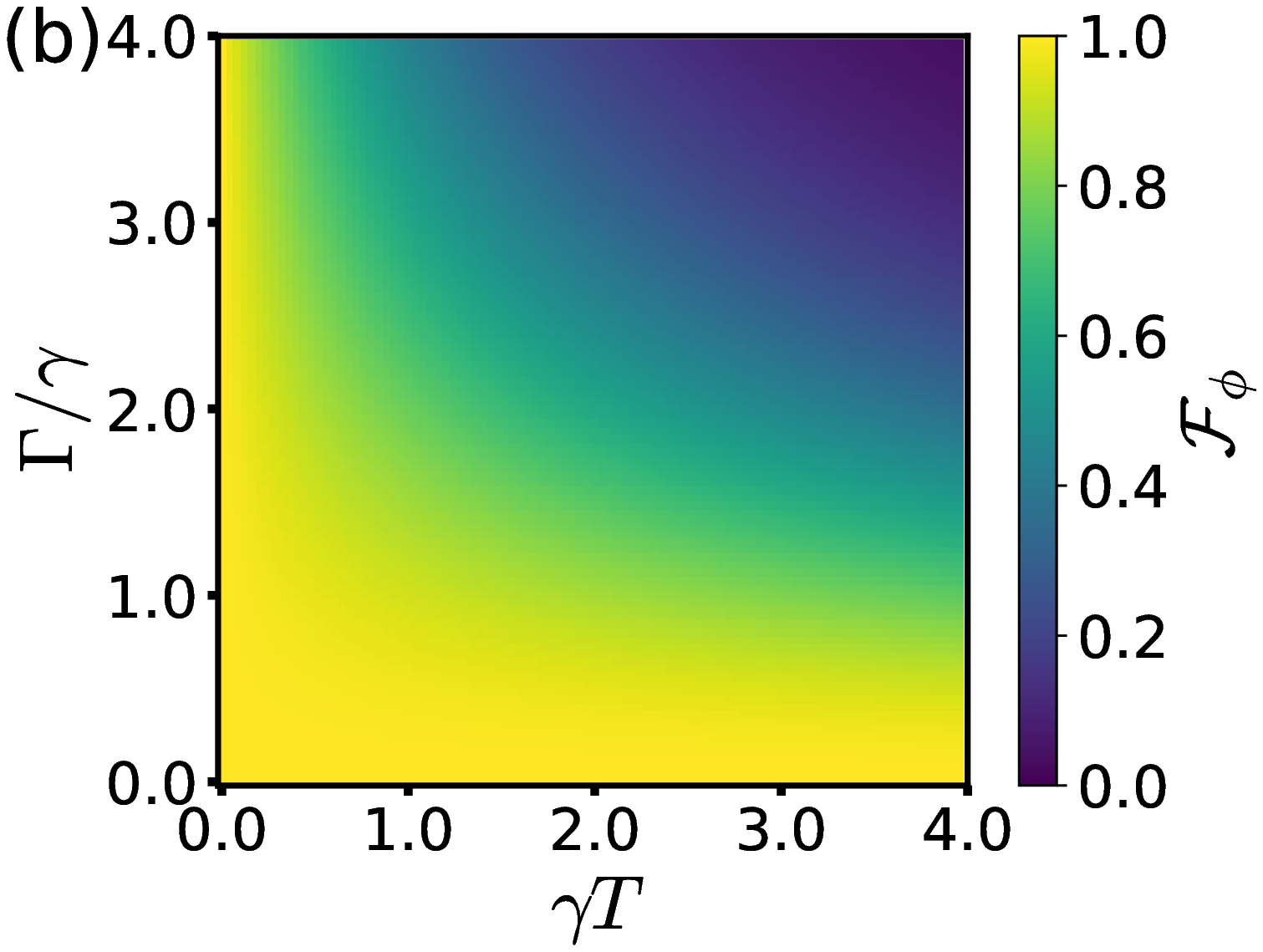}
\caption{Landscape of the gate-fidelity $\mathcal{F}_{\phi}$ under the stochastic $\dot{\phi}$ in the parameter space of the strength-memory ratio $\Gamma/\gamma$ and the running period $\gamma T$. In (a) and (b), the noise-free parameter $\theta$ is set as $\theta(t)=\omega t$ and $\theta(t)=\omega t-\sin(\omega t)$, respectively. } \label{FigFidelity}
\end{figure}

Now we consider that the imperfect control over the phase $\phi$ in the presence of random noise associated with its time-derivative $\dot{\phi}$, which leads to $\phi(T)\rightarrow\phi(T)+\Delta_{\phi}(T)$ with $\Delta_{\phi}(T)=\int_0^Tdt\delta_{\phi}(t)$. Then in the rotating frame, the nonideal Hamiltonian for the accelerated phase gate becomes,
\begin{equation}\label{Hnoisephi}
\tilde{H}_z(t)=\frac{1}{2}\left\{B_0-\left[\dot{\phi}+\delta_{\phi}(t)\right]
\cos\theta\right\}\sigma_z.
\end{equation}
Using Eqs.~(\ref{Fidelity}), (\ref{plusrhoTminus}), and (\ref{Hnoisephi}), it is straightforward to find for $\theta=\omega t$ that
\begin{equation}\label{Fidelityphi}
\begin{aligned}
&\mathcal{F}_{\phi}=e^{-\frac{1}{2}\int_0^Tdt\int_0^Tds \left[1-\cos\theta(t)\right]\left[1-\cos\theta(s)\right]C(t-s)}\\
&=\exp\Big\{-\frac{\Gamma}{4\gamma(4\pi^2+\gamma^2T^2)^2}\big[32\pi^4e^{-\gamma T}\\ &+32\pi^4\left(-1+\gamma T\right)+20\pi^2\left(\gamma T\right)^3+3\left(\gamma T\right)^5\big]\Big\},
\end{aligned}
\end{equation}
Similar to $\mathcal{F}_B$ in Eq.~(\ref{FidelityB}), the fidelity $\mathcal{F}_{\phi}$ in Eq.~(\ref{Fidelityphi}) is also a function of dimensionless parameters $\gamma T$ and $\Gamma/\gamma$. An interesting observation is that under the short time limit, i.e., $\gamma T\ll1$, we have $\mathcal{F}_{\phi}\approx\exp(-\Gamma\gamma T^2/4)$, the same as Eq.~(\ref{ApproximatedFIdelityB}). Then we obtain the same decoherence timescale $T_2$ as in Eq.~(\ref{T2B}). One can hardly distinguish Fig.~\ref{FigFidelityB} and Fig.~\ref{FigFidelity}(a). It is found when $\Gamma/\gamma\leq1.0$, $\mathcal{F}_{\phi}$ is over $0.90$ at $\gamma T=0.69$.

Equation~(\ref{Fidelityphi}) is however reminiscent of an improved scheme in constructing the quantum phase-gate by virtue of the time-dependence of $\theta$. Rather than a constant $\theta$ used in literature~\cite{AAphaseQuantumGate2,AAphaseQuantumGate}, one can reduce the magnitude of the integrand in Eq.~(\ref{Fidelityphi}) by selecting a proper and experimentally accessible $\theta(t)$, provided that the boundary condition is satisfied. For the exponent of the fidelity in Eq.~(\ref{Fidelityphi}), the main contribution to the integral is around $t-s\approx0$ according to the exponential decay of the correlation function $C(t-s)$ [see e.g., Eq.~(\ref{Statistic}), and the monotonic or asymptotic decay behavior is popular for all the stationary Gaussian noises]. Then we have $\mathcal{F}_{\phi}\approx\exp\{-1/2[\int_0^Tdt(1-\cos\theta(t))]^2\}$. The fidelity can thus be enhanced to a certain extent by reducing the magnitude of $\int_0^Tdt[1-\cos\theta(t)]$. For example, one can replace $\theta(t)=\omega t$ with $\theta(t)=\omega t-\sin(\omega t)$ that holds the same boundary condition. The induced improvement in gate-fidelity can be observed in Fig.~\ref{FigFidelity}(b). In contrast to Fig.~\ref{FigFidelity}(a), $\mathcal{F}_{\phi}$ is over $0.90$ with $\Gamma/\gamma=3.1$ when $\gamma T=0.69$. And when $\Gamma/\gamma=1.0$, the same high-level fidelity can be sustained until $\gamma T=1.24$.

\section{Suppress noise by dynamical decoupling}\label{secDD}

The noise analysis over both $\delta_B$ and $\delta_{\phi}$ renders the same second-order behavior in the running time under the short-time limit, as suggested by Eqs.~(\ref{ApproximatedFIdelityB}) and (\ref{T2B}). The gate-fidelity decay induced by the classical noise on the control parameters can be suppressed by a sequence of dynamical decoupling. As a developed technique, dynamical decoupling can extend the coherence time in many experiments. It has been generalized into various sequences of pulse to neutralize the influence of the environmental noises~\cite{ReviewsofDD}. Spin echo (SE)~\cite{SpinEcho} presents the simplest yet the original form in these pulse sequences, which consists of only one $\pi$-pulse in the middle of the time evolution besides another one performed in the end. Based on SE, the Carr-Purcell-Meiboom-Gill sequence (CPMG)~\cite{CPMG1,CPMG2,MultipleCPMG} employs two or more $\pi$-pulses. In general, the $n$-pulse version of CPMG~\cite{MultipleCPMG} can be described by a sequence of $t_k=(k-1/2)T/n$, $k=1,2,\cdots,n$, that is obtained in the frequency domain. When the noise spectrum has a hard cutoff~\cite{UDD2}, i.e., $S(\omega)\sim\omega\Theta(\omega_c-\omega)$, where $\Theta(x)$ is the Heaviside step function [$\Theta(x)=1$ when $x\ge0$ and $\Theta(x)=0$ when $x<0$] or has an exponential-decay cutoff, i.e., $S(\omega)\sim\exp(-\omega/\omega_c)$, the Uhrig dynamical-decoupling (UDD) sequence~\cite{UDD1} is shown to be the most efficient scheme. It can reduce the decoherence rate down to the $n$th order of the running time by using $n$ non-periodical pulses. In this section, we focus on the noise spectrum of the magnetic field $\delta_B(t)$ following the inverse-quadratic power-law. It is shown that CPMG is the optimized choice instead of UDD. The analysis can be straightforwardly extended to the noisy $\dot{\phi}$.

\subsection{Spin echo on geometric phase}\label{secSE}

\begin{figure}[htbp]
\includegraphics[width=0.4\textwidth]{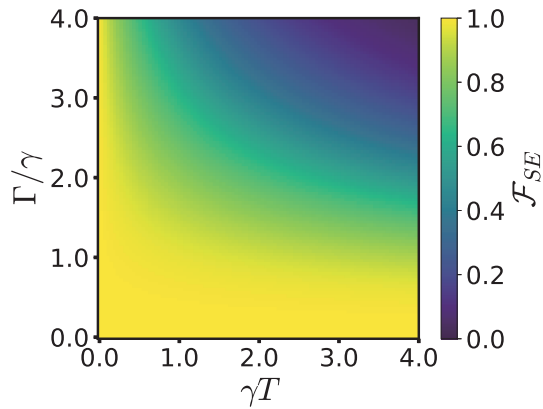}
\caption{Landscape of the spin-echo gate-fidelity $\mathcal{F}_{\rm SE}$ under the stochastic $B_0$ in the parameter space of the strength-memory ratio $\Gamma/\gamma$ and the running period $\gamma T$.}\label{FigSE}
\end{figure}

We choose SE as a warm-up example to illustrate the DD effect on the geometric phase. The other pulse sequences can be analysed in a similar way. To focus on the sequence itself, the pulses are considered to be ideal, i.e., instantaneous and with no error~\cite{UDD1,MultipleCPMG}. The SE scheme is a concatenated process of two piecewise segments, i.e., $0\rightarrow t_f$ and $t_f\rightarrow T$, where $t_f=T/2$. A $\pi$-pulse is inserted at the moment $t_f$ to switch the signs of the eigenstates. Another $\pi$-pulse is imposed on the system in the end to complete a cyclic process. Starting from the initial state $|\psi(0)\rangle=|+\rangle$ and denoting the instantaneous eigenstates as $|\pm(t)\rangle$, the whole process can be described by
\begin{equation}
\begin{aligned}
&|\psi(0)\rangle=|+\rangle\rightarrow|\psi(t_f-0^+)\rangle = e^{i\bar{\gamma}_1}\left|+(t_f)\right\rangle\xrightarrow{\pi\,pulse}\\
&|\psi(t_f+0^+)\rangle = e^{i\bar{\gamma}_1}\left|-(t_f)\right\rangle\rightarrow\\
&\left|\psi(T-0^+)\right\rangle = e^{i(\bar{\gamma}_1+\bar{\gamma}_2)}\left|-(T)\right\rangle\xrightarrow{\pi\,pulse}\\
&|\psi(T)\rangle= e^{i(\bar{\gamma}_1+\bar{\gamma}_2)} \left|+(T)\right\rangle=e^{i(\pi+\bar{\gamma}_1+\bar{\gamma}_2)} |+\rangle,
\end{aligned}
\end{equation}
where the boundary condition has been applied in the last equation. $\bar{\gamma}_1$ and $\bar{\gamma}_2$ represent the quantum phases generated in the first and the second segments, respectively. According to Eqs.~(\ref{idealUT}), (\ref{Totalphase}), and (\ref{HnoiseB}), these two nonideal phases turn out to be
\begin{equation}\label{bargamma}
\begin{aligned}
&\bar{\gamma}_1=-\frac{1}{2}\int_0^{T/2}dt\left[B_0+\delta_B(t)-\dot{\phi}'(\cos\theta'-1)\right],\\
&\bar{\gamma}_2=\frac{1}{2}\int_{T/2}^{T}dt\left[B_0+\delta_B(t)-\dot{\phi}'(\cos\theta'-1)\right],
\end{aligned}
\end{equation}
where $\phi'=\phi$ and $\theta'=\theta$ during the first-half period $t\in(0,t_f)$ and $\phi'=-\phi$ and $\theta'=\theta$ during the second-half period $t\in(t_f,T)$ due to the spin-echo effect. By Eq.~(\ref{bargamma}), the ideal dynamical phase proportional to $B_0$ vanishes in the end of the running while the geometric phase is accumulated to realize a desired transformation and it holds the same form as in Eq.~(\ref{AAphase}). When the initial state starts from $|\psi(0)\rangle=|-\rangle$, the signs of those phases in Eq.~(\ref{bargamma}) are reversed and the SE scheme still works. In any loop, however, the dynamical phase associated with the accumulation of the stochastic noise $\delta_B(t)$ during the two consecutive time-integrals has not been completely eliminated.

Then we measure the gate-fidelity in the presence of the random magnetic field under the spin echo scheme. The ratio of the off-diagonal elements of the density matrix in Eq.~(\ref{Fidelity}) is found to be
\begin{equation}
\frac{\langle+|\rho(T)|-\rangle}{\langle+|\psi(T)\rangle\langle\psi(T)|-\rangle} = e^{-i\left[\int_0^{T/2}dt\delta_B(t)-\int_{T/2}^Tdt\delta_B(t)\right]}.
\end{equation}
On ensemble average, we have
\begin{equation}\label{FidelitySE}
\begin{aligned}
&\mathcal{F}_{\rm SE}=M\left[\exp\left(-i\int_0^{T/2}dt\delta_B(t)+i\int_{T/2}^Tdt\delta_B(t)\right)\right] \\
&=\exp\left[-\frac{\Gamma}{2\gamma}\left(\gamma T-e^{-\gamma T}-3+4e^{-\gamma T/2}\right)\right],
\end{aligned}
\end{equation}
as plotted in Fig.~\ref{FigSE}. In contrast to Fig.~\ref{FigFidelityB}, one can find that under the SE scheme, the high-level regime of the gate-fidelity is significantly enlarged. When $\Gamma/\gamma=2.0$, $\mathcal{F}_{\rm SE}$ can be maintained over $0.98$ for $\gamma T=1.0$. In the short-time limit $\gamma T\ll1$, we have
\begin{equation}
\mathcal{F}_{\rm SE}\approx\exp\left(-\frac{\Gamma\gamma^2}{24}T^3\right).
\end{equation}
In comparison to the quadratic dependence in Eq.~(\ref{ApproximatedFIdelityB}), the exponent function is reduced to the third-order of the running time of the geometric gate by the spin echo effect. It is shown in Appendix~\ref{appendix} that for the inversely quadratic power-law noise spectrum, one can further reduce the decay coefficient by more DD operations, yet cannot reduce the exponent function to more higher orders of the running interval $T$.

\subsection{CPMG on geometric phase}\label{secCPMG}

Generally, we apply $n$ $\pi$-pulses into the running period $(0, T)$ to suppress the geometric-phase error. The whole process is divided into $n+1$ segments described by $0\rightarrow t_1$, $t_1\rightarrow t_2$, $\cdots$, $t_n\rightarrow T$. We are desired to seek an optimized sequence of the DD-operation moments $t_k$. Similar to Eq.~(\ref{FidelitySE}), the gate-fidelity can be represented by
\begin{equation}\label{FidelityDD}
\mathcal{F}_n=M\left[e^{-i\int_0^Tdt\delta_B(t)f(T;t)}\right],
\end{equation}
where $f(T;t)=\sum_{k=0}^n(-1)^k\Theta(t_{k+1}-t)\Theta(t-t_k)$ with $t_0=0$ and $t_{n+1}=T$, indicating the $\pi$-pulse-induced sign change of the quantum phase. Moreover, to hold the same geometric phase as that in Eq.~(\ref{AAphase}) for the FID process, the phase parameters $\phi'$, $\theta'$ in all the segments are set as $\phi'=f(T;t)\phi$ and $\theta'=\theta$.

The gate-fidelity in Eq.~(\ref{FidelityDD}) can be represented by $\mathcal{F}_n\equiv e^{-\chi(T)}$~\cite{PulseSequenceAnalysis2,PulseSequenceAnalysis1}, where the decay function $\chi(T)$ is obtained by the noise spectrum $S(\omega)$ and $\tilde{f}(\omega,T)=\int_0^Tdte^{-i\omega t}f(T;t)$ [the Fourier transform of $f(T;t)$],
\begin{equation}\label{chi}
\begin{aligned}
&\chi(T)=\frac{1}{2}\int_{-\infty}^\infty\frac{d\omega}{2\pi}|\tilde{f}(\omega,T)|^2S(\omega) \\
&=\frac{1}{2}\int_{-\infty}^\infty\frac{d\omega}{\pi}\frac{F(\omega T)}{\omega^2}S(\omega).
\end{aligned}
\end{equation}
Note in the second line of Eq.~(\ref{chi}), a filter function $F(\omega T)\equiv|\omega\tilde{f}(\omega,T)|^2/2$ appears to measure the effects of the pulse sequence. Using the Taylor expansion around $\omega T=0$, we have
\begin{equation}
\begin{aligned}
\chi(T)&=\int_{-\infty}^\infty\frac{d\omega}{2\pi\omega^2}\Big[F(0)+F'(0)\omega T+F''(0)\frac{(\omega T)^2}{2!} \\ &+F^{(3)}(0)\frac{(\omega T)^3}{3!}+\cdots+\Big]S(\omega).
\end{aligned}
\end{equation}
Under the general noise with a power-law spectrum $S(\omega)\sim 1/\omega^m$, $m\geq2$, the high-order terms $F^{(n)}(0)$ with $n-m\geq 1$ would lead to the nonconvergent integration $\int_{-\infty}^\infty d\omega\omega^{n-2-m}$. In contrast, if $S(\omega)$ has a hard cutoff at an upper-bound frequency $\omega_c$ or has an exponential decay with frequency, such as $S(\omega)\sim\exp(-\omega/\omega_c)$, then the convergency could be hold at any order. Without frequency cutoff, the UDD scheme~\cite{UDD1} does not necessarily supply an optimized sequence for polynomial spectrum.

Some existing works~\cite{PulseSequenceAnalysis1,PulseSequenceAnalysis2} about DD have addressed the noise $\delta_B(t)$ with the statistical properties in Eq.~(\ref{Statistic}). Here we provide an alternative way in Appendix~\ref{appendix} to optimize the pulse sequence in the time domain for the quadratic power-law spectrum. We find the most efficient sequence can be analytically described by
\begin{equation}\label{TimeInterval}
t_k=\frac{(k-1/2)}{n}T, \quad k=1, 2, \cdots, n,
\end{equation}
which is exactly the CPMG$n$ sequence with $n\ge2$. When $n=1$, it reduces to the SE sequence. When $n=2$, it is also coincident to the UDD sequence.

\begin{figure}[htbp]
\includegraphics[width=0.4\textwidth]{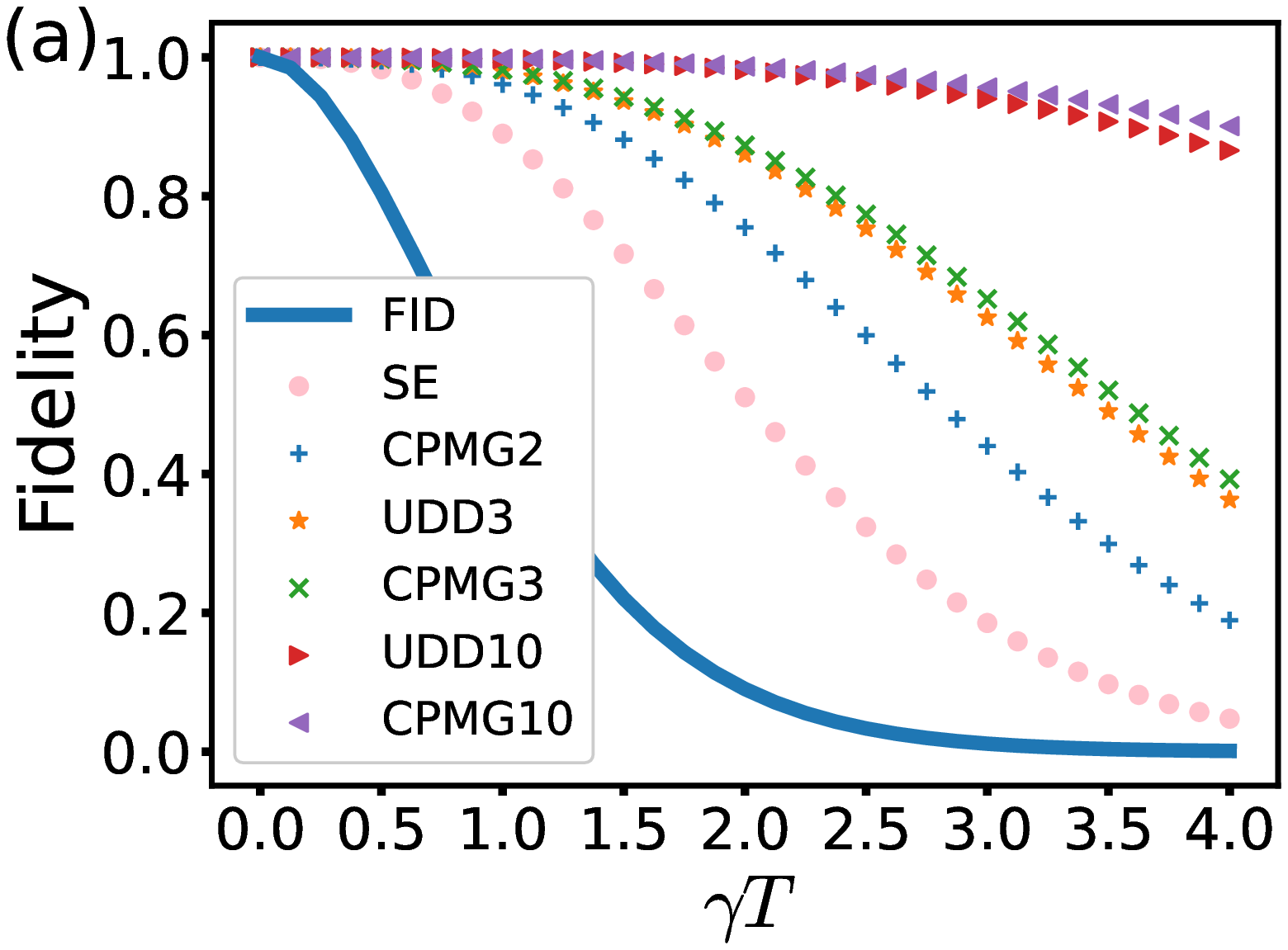}
\includegraphics[width=0.4\textwidth]{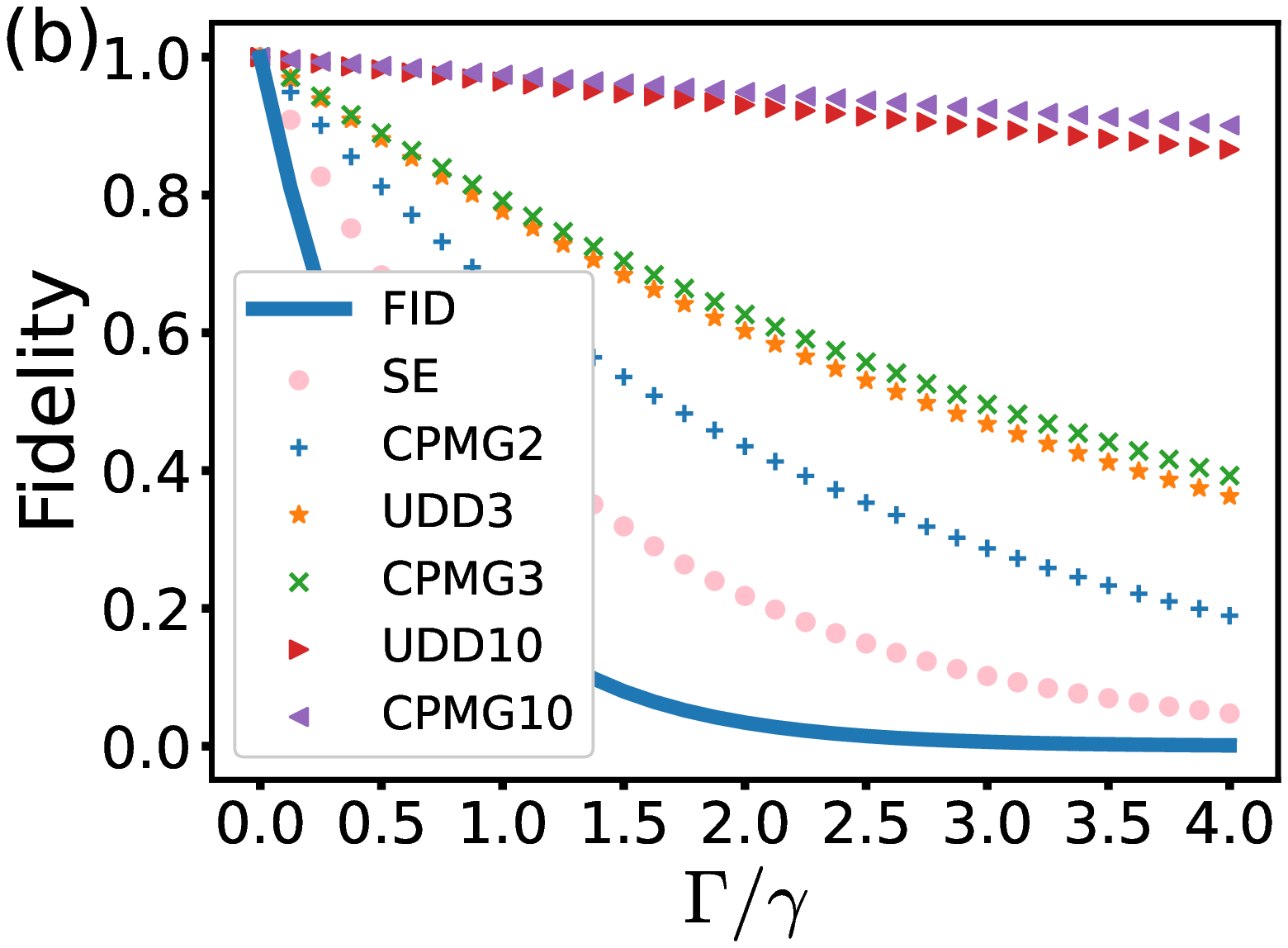}
\caption{The gate-fidelity under various DD sequences (SE, CPMG$n$, and UDD$n$) in the presence of the Gaussian noise following the inverse-quadratic power-law spectrum. (a) the gate dynamics versus $\gamma T$ when $\Gamma/\gamma=4$; (b) the gate dependence on $\Gamma/\gamma$ when $\gamma T=4$. } \label{FigSequenceCompare}
\end{figure}

In Figs.~\ref{FigSequenceCompare}(a) and (b), we show the gate-fidelities under various DD-sequence in terms of the dynamics and their dependence on the memory parameter, respectively. The solid line, circle-dotted line, plus-dotted line, star-dotted line, cross-dotted line, right-triangle-dotted line, and left-triangle-dotted line represent the fidelities under FID, SE, CPMG$2$, UDD$3$, CPMG$3$, UDD$10$, and CPMG$10$, respectively, where $n$ denotes the number of $\pi$-pulses. In both Figs.~\ref{FigSequenceCompare}(a) and \ref{FigSequenceCompare}(b), the performance of the geometric phase gate is steadily improved by inserting more periodical $\pi$-pulses into the running period. We also demonstrate that CPMG$n$ outperforms UDD$n$ when $n\geq3$. It is found when $\Gamma/\gamma=4.0$ and $\gamma T=4.0$, the gate-fidelity can be maintained over $0.90$ by CPMG$10$. By Eqs.~(\ref{TimeInterval}) and (\ref{chi}), it is found that up to the leading-order contribution the gate-fidelity becomes
\begin{equation}\label{chiApproximation}
\mathcal{F}_n\approx\exp\left[-\frac{\Gamma}{24n^2\gamma}\left(\gamma T\right)^3\right].
\end{equation}
As for the scaling of the decoherence time $T_2$ over the pulse number $n$, we have
\begin{equation}
T_2=\left(\frac{24}{\Gamma\gamma^2}\right)^{1/3}n^{2/3}.
\end{equation}
The power-law with $n^{2/3}$ implies that more DD pulses are required to enhance the coherence time of our geometric phase gate.

\section{Discussion and conclusion}\label{secConclusion}

The analysis over the classical noise in this work is mainly based on the Gaussian approximation. Then the two-point correlation function $C(t-s)$ is sufficient to determine its impact on the decoherence behavior in the time evolution. As pointed out in literature~\cite{PulseSequenceAnalysis1,PulseSequenceAnalysis2}, random telegraph noise in the weak noise regime can be described by Eq.~(\ref{Statistic}) and the power-law noise can be described by a linear combination of the similar correlation functions with various parameters. Under the non-Gaussian assumption, the noise effect also involves with multiple-point correlation function. It is thus not convenient to analytically obtain $\mathcal{F}_B$ in Eq.~(\ref{FidelityB}), $\mathcal{F}_{\phi}$ in Eq.~(\ref{Fidelityphi}), and $\mathcal{F}_{SE}$ in Eq.~(\ref{FidelitySE}) to witness the effect of a non-Markovian noise on the gate-fidelity. Although the non-Gaussian noise would reduce the coherence time in FID, however, it has been shown to be more suppressed under CPMG rather than UDD~\cite{PulseSequenceAnalysis1}. The non-Gaussian noise is therefore negligible when applying our dynamical-decoupling scheme, that is proved to be a CPMG, into the accelerated phase-gate construction.

In summary, we construct a quick-and-faith geometric phase gate by combining the transitionless-quantum-driving approach and the dynamical decoupling control into the Berry phase. The former is used to shorten the running time as required by the adiabatic passage and the latter is used to neutralize the classical noise on the control parameters. Our proposal is based on a semi-classical Rabi model under a parametric driving. Using a modified transformation to hold the first-order contribution from the counter-rotating interaction, our effective Hamiltonian is superior to that under the conventional rotating-wave approximation and adapts to a strong driving beyond the dispersive regime. We analysis the gate-fidelity under the random fluctuation or noise on the effective magnetic field and the control phase. In the time domain, we find that the CPMG$n$ sequence is the most efficient DD scheme against the inversely quadratic power-law noise to maintain a high-level fidelity and we obtain the scaling behavior of the decoherence time $T_2$ with respect to the DD operation number $n$. Our investigation provides a systematic estimation over the errors in constructing the Berry-phase gate caused by the classical noise. It is useful in optimizing the performance of quantum gates under Gaussian noise following a power-law spectrum.

In addition, combining the noninstantaneous DD pulses with the approximated TQD approaches~\cite{approTQD1,approTQD2} would be a near-future target. The phase-gate fidelity is supposed to be more robust when the counterdiabatic operation could be performed along the same direction as the DD pulses.

\section*{Acknowledgments}

We acknowledge financial support from the National Science Foundation of China (Grants No. 11974311 and No. U1801661).

\appendix

\section{Optimized sequence for Gaussian noise of power-law spectrum}\label{appendix}

This appendix contributes to optimizing DD sequence to cancel the classical Gaussian noise on the magnetic-field intensity. We provide a proof in the time domain rather than that in the frequency domain~\cite{PulseSequenceAnalysis2}. The derivation starts from the gate-fidelity under $n$ pulses in Eq.~(\ref{FidelityDD}). Using the correlation function in Eq.~(\ref{Statistic}), we have
\begin{equation}
\begin{aligned}
&\mathcal{F}_n\left(\{t_k, k=1,2,\cdots, n\}\right)=M\left[e^{-i\int_0^Tdt\delta_B(t)f(T;t)}\right]\\
&=\exp\left[-\frac{1}{2}\int_0^Tdt\int_0^TdsC(t-s)f(T;t)f(T;s)\right].
\end{aligned}
\end{equation}
With the dimensionless parameters $x\equiv\gamma T$, $y\equiv\Gamma/\gamma$ and $\mu_k\equiv t_k/T$, the decay function $\chi(x,y)\equiv-\ln\mathcal{F}_n$ can be expressed by
\begin{equation}
\begin{aligned}
&\chi(x,y)=\frac{y}{2}\Big[x-1+(-1)^ne^{-x}\\
&-2\sum_{k=0}^{n+1}\sum_{j=1}^{n}(-1)^{|k-j|}e^{-|\mu_k-\mu_j|x}\Big],
\end{aligned}
\end{equation}
where $\mu_0=0$ and $\mu_{n+1}=1$. Note expanding $\chi(x,y)$ about $x$ is equivalent to expanding it about $T$. It is obtained that
\begin{equation}
\chi(x,y)=\frac{y}{2}\left(C_0+C_1x+C_2x^2+C_3x^3+\cdots\right),
\end{equation}
where the coefficients of the first few orders are
\begin{subequations}
\begin{equation}\label{0order}
C_0=-1+(-1)^n-2\sum_{k=0}^{n+1}\sum_{j=1}^n(-1)^{|k-j|},
\end{equation}
\begin{equation}\label{1order}
C_1=1-(-1)^n+2\sum_{k=0}^{n+1}\sum_{j=1}^n(-1)^{|k-j|}|\mu_k-\mu_j|,
\end{equation}
\begin{equation}\label{2order}
C_2=\frac{1}{2}(-1)^n-\sum_{k=0}^{n+1}\sum_{j=1}^n(-1)^{|k-j|}|\mu_k-\mu_j|^2,
\end{equation}
\begin{equation}\label{3order}
C_3=\frac{1}{6}(-1)^{n+1}+\frac{1}{3}\sum_{k=0}^{n+1}\sum_{j=1}^n(-1)^{|k-j|}|\mu_k-\mu_j|^3.
\end{equation}
\end{subequations}
It is straightforward to verify that both the zero-order and the first-order coefficients $C_0$ and $C_1$ are exactly $0$, irrespective to the sequence arrangement $\{\mu_k\}$.

When $n$ is an even integer, the summation in the second-order coefficient $C_2$ can be decomposed into three terms, being formally relevant to $\mu_k^2$, $-2\mu_k\mu_j$, and $\mu_j^2$, respectively. By virtue of $\mu_0=0$ and $\mu_{n+1}=1$, the $\mu_k^2$ or $\mu_j^2$ term is proved to vanish by
\begin{equation}
\sum_{k=0}^{n+1}\sum_{j=1}^n(-1)^{|k-j|}\mu_k^2=\sum_{k=0}^{n+1}\mu_k^2\sum_{j=1}^n(-1)^{|k-j|}=0.
\end{equation}
Equation~(\ref{2order}) can then be reduced to
\begin{equation}\label{dynphasecondition}
\begin{aligned}
& C_2=\frac{1}{2}+2\sum_{k=0}^{n+1}\sum_{j=1}^n(-1)^{|k-j|}\mu_k\mu_j\\
&=\frac{1}{2}+2\left[\sum_{k=1}^n(-1)^{k-1}\mu_k\right]^2+2\sum_{k=1}^{n}(-1)^{k-1}\mu_k\\
&=\frac{1}{2}\left[1+\sum_{k=1}^n(-1)^{k-1}2\mu_k\right]^2.
\end{aligned}
\end{equation}

When $n$ is odd, the summations about $\mu_k^2$ and $\mu_j^2$ in $C_2$ are simplified to
\begin{equation}
\begin{aligned}
&\sum_{k=0}^{n+1}\mu_k^2\sum_{j=1}^n(-1)^{|k-j|} + \sum_{j=1}^n\mu_j^2\sum_{k=0}^{n+1}(-1)^{|k-j|}\\
&=\sum_{k=0}^{n+1}\mu_k^2(-1)^{|k-1|}+\sum_{j=1}^n\mu_j^2(-1)^j\\
&=\mu_0^2(-1)^1+\mu_{n+1}^2(-1)^n=-1.
\end{aligned}
\end{equation}
Then Eq.~(\ref{2order}) can be reduced to
\begin{equation}\label{oddcondition}
\begin{aligned}
& C_2=-\frac{1}{2}-(-1)+2\sum_{k=0}^{n+1}\sum_{j=1}^n(-1)^{|k-j|}\mu_k\mu_j\\
&=\frac{1}{2}+2\left[\sum_{k=1}^n(-1)^{k-1}\mu_k\right]^2-2\sum_{k=1}^{n}(-1)^{k-1}\mu_k\\
&=\frac{1}{2}\left[-1+\sum_{k=1}^n(-1)^{k-1}2\mu_k\right]^2.
\end{aligned}
\end{equation}
Summarizing Eqs.~(\ref{dynphasecondition}) and (\ref{oddcondition}), the condition of $C_2=0$ can be written as
\begin{equation}\label{mun}
(-1)^n+2\sum_{k=1}^n(-1)^{k-1}\mu_k=0.
\end{equation}
It has a clear physical indication or consequence that after applying those $\pi$-pulses at $\mu_k$, the dynamical phase in the absence of noise is thus exactly eliminated to leave a geometric transformation.

To find the optimized sequence, we now study the extreme-value condition $\partial_{\mu_k}C_3=0$, $k=1,2,\cdots,n$, under the constrains that $0<\mu_1<\mu_2<\cdots<\mu_n<1$. By virtue of Eq.~(\ref{mun}), we have
\begin{equation}
\mu_n=\sum_{k=1}^{n-1}(-1)^{n-k-1}\mu_k+\frac{1}{2}.
\end{equation}
With Eq.~(\ref{3order}), we have for $k=n-1$ that
\begin{equation}
\begin{aligned}
&\frac{\partial C_3}{\partial\mu_{n-1}}=(\mu_{n-1}-\mu_n)\Big[2-\mu_{n-1}-\mu_n\\
&+(-1)^{n-1}(\mu_{n-1}+\mu_n) \\
&+2\sum_{k=1}^{n-2}(-1)^{n-k-1}(\mu_{n-1}+\mu_n-2\mu_k)\Big]=0
\end{aligned}
\end{equation}
For $n$ is either even or odd, it gives rise to
\begin{equation}\label{mun-1}
\mu_{n-1}=\frac{3}{2}\sum_{k=1}^{n-2}(-1)^{n-k-2}\mu_k+\frac{1}{4}.
\end{equation}
By a similar derivation with decreasing $k$, we can obtain a general formula
\begin{equation}\label{generalmu}
\mu_{n-j}=\frac{2j+1}{j+1}\sum_{k=1}^{n-j-1}(-1)^{n-k-j-1}\mu_k+\frac{1}{2j+2}.
\end{equation}
In the end, it is found that $\mu_1=1/(2n)$. Then by iteratively using Eq.~(\ref{generalmu}), we can find the solution is exactly the CPMG$n$, i.e., $\mu_k=(k-1/2)/n$. And the minimum-value of the third-order coefficient $C_3$ turns out to be $x^3/(12n^2)$, which corresponds to the decay behavior in Eq.~(\ref{chiApproximation}) of the main text. The result also demonstrates that the $C_3$ cannot be completely eliminated under arbitrary choice of $\mu_k$ or $t_k$.

\bibliographystyle{apsrevlong}
\bibliography{reference}

\begin{thebibliography}{51}%
\makeatletter
\providecommand \@ifxundefined [1]{%
 \@ifx{#1\undefined}
}%
\providecommand \@ifnum [1]{%
 \ifnum #1\expandafter \@firstoftwo
 \else \expandafter \@secondoftwo
 \fi
}%
\providecommand \@ifx [1]{%
 \ifx #1\expandafter \@firstoftwo
 \else \expandafter \@secondoftwo
 \fi
}%
\providecommand \natexlab [1]{#1}%
\providecommand \enquote  [1]{``#1''}%
\providecommand \bibnamefont  [1]{#1}%
\providecommand \bibfnamefont [1]{#1}%
\providecommand \citenamefont [1]{#1}%
\providecommand \href@noop [0]{\@secondoftwo}%
\providecommand \href [0]{\begingroup \@sanitize@url \@href}%
\providecommand \@href[1]{\@@startlink{#1}\@@href}%
\providecommand \@@href[1]{\endgroup#1\@@endlink}%
\providecommand \@sanitize@url [0]{\catcode `\\12\catcode `\$12\catcode
  `\&12\catcode `\#12\catcode `\^12\catcode `\_12\catcode `\%12\relax}%
\providecommand \@@startlink[1]{}%
\providecommand \@@endlink[0]{}%
\providecommand \url  [0]{\begingroup\@sanitize@url \@url }%
\providecommand \@url [1]{\endgroup\@href {#1}{\urlprefix }}%
\providecommand \urlprefix  [0]{URL }%
\providecommand \Eprint [0]{\href }%
\providecommand \doibase [0]{http://dx.doi.org/}%
\providecommand \selectlanguage [0]{\@gobble}%
\providecommand \bibinfo  [0]{\@secondoftwo}%
\providecommand \bibfield  [0]{\@secondoftwo}%
\providecommand \translation [1]{[#1]}%
\providecommand \BibitemOpen [0]{}%
\providecommand \bibitemStop [0]{}%
\providecommand \bibitemNoStop [0]{.\EOS\space}%
\providecommand \EOS [0]{\spacefactor3000\relax}%
\providecommand \BibitemShut  [1]{\csname bibitem#1\endcsname}%
\let\auto@bib@innerbib\@empty
%</preamble>
\bibitem [{\citenamefont {Shor}(1997)}]{ShorAlgorithm}%
  \BibitemOpen
  \bibfield  {author} {\bibinfo {author} {\bibfnamefont {P.~W.}\ \bibnamefont
  {Shor}},\ }\bibfield  {title} {\emph {\bibinfo {title} {Polynomial-time
  algorithms for prime factorization and discrete logarithms on a quantum
  computer},\ }}\href {\doibase 10.1137/S0097539795293172} {\bibfield
  {journal} {\bibinfo  {journal} {SIAM J. Comput.}\ }\textbf {\bibinfo {volume}
  {26}},\ \bibinfo {pages} {1484} (\bibinfo {year} {1997})}\BibitemShut
  {NoStop}%
\bibitem [{\citenamefont {Grover}(1997)}]{GroverAlgorithm}%
  \BibitemOpen
  \bibfield  {author} {\bibinfo {author} {\bibfnamefont {L.~K.}\ \bibnamefont
  {Grover}},\ }\bibfield  {title} {\emph {\bibinfo {title} {Quantum mechanics
  helps in searching for a needle in a haystack},\ }}\href {\doibase
  10.1103/PhysRevLett.79.325} {\bibfield  {journal} {\bibinfo  {journal} {Phys.
  Rev. Lett.}\ }\textbf {\bibinfo {volume} {79}},\ \bibinfo {pages} {325}
  (\bibinfo {year} {1997})}\BibitemShut {NoStop}%
\bibitem [{\citenamefont {Babbush}\ \emph {et~al.}(2014)\citenamefont
  {Babbush}, \citenamefont {Love},\ and\ \citenamefont
  {Aspuru-Guzik}}]{QuantumSimulation1}%
  \BibitemOpen
  \bibfield  {author} {\bibinfo {author} {\bibfnamefont {R.}~\bibnamefont
  {Babbush}}, \bibinfo {author} {\bibfnamefont {P.~J.}\ \bibnamefont {Love}}, \
  and\ \bibinfo {author} {\bibfnamefont {A.}~\bibnamefont {Aspuru-Guzik}},\
  }\bibfield  {title} {\emph {\bibinfo {title} {Adiabatic quantum simulation of
  quantum chemistry},\ }}\href {\doibase https://doi.org/10.1038/srep06603}
  {\bibfield  {journal} {\bibinfo  {journal} {Sci Rep}\ }\textbf {\bibinfo
  {volume} {4}},\ \bibinfo {pages} {1} (\bibinfo {year} {2014})}\BibitemShut
  {NoStop}%
\bibitem [{\citenamefont {Gerritsma}\ \emph {et~al.}(2010)\citenamefont
  {Gerritsma}, \citenamefont {Kirchmair}, \citenamefont {Z{\"a}hringer},
  \citenamefont {Solano}, \citenamefont {Blatt},\ and\ \citenamefont
  {Roos}}]{QuantumSimulation2}%
  \BibitemOpen
  \bibfield  {author} {\bibinfo {author} {\bibfnamefont {R.}~\bibnamefont
  {Gerritsma}}, \bibinfo {author} {\bibfnamefont {G.}~\bibnamefont
  {Kirchmair}}, \bibinfo {author} {\bibfnamefont {F.}~\bibnamefont
  {Z{\"a}hringer}}, \bibinfo {author} {\bibfnamefont {E.}~\bibnamefont
  {Solano}}, \bibinfo {author} {\bibfnamefont {R.}~\bibnamefont {Blatt}}, \
  and\ \bibinfo {author} {\bibfnamefont {C.}~\bibnamefont {Roos}},\ }\bibfield
  {title} {\emph {\bibinfo {title} {Quantum simulation of the dirac equation},\
  }}\href {\doibase https://doi.org/10.1038/nature08688} {\bibfield  {journal}
  {\bibinfo  {journal} {Nature}\ }\textbf {\bibinfo {volume} {463}},\ \bibinfo
  {pages} {68} (\bibinfo {year} {2010})}\BibitemShut {NoStop}%
\bibitem [{\citenamefont {Kivlichan}\ \emph {et~al.}(2018)\citenamefont
  {Kivlichan}, \citenamefont {McClean}, \citenamefont {Wiebe}, \citenamefont
  {Gidney}, \citenamefont {Aspuru-Guzik}, \citenamefont {Chan},\ and\
  \citenamefont {Babbush}}]{QuantumSimulation3}%
  \BibitemOpen
  \bibfield  {author} {\bibinfo {author} {\bibfnamefont {I.~D.}\ \bibnamefont
  {Kivlichan}}, \bibinfo {author} {\bibfnamefont {J.}~\bibnamefont {McClean}},
  \bibinfo {author} {\bibfnamefont {N.}~\bibnamefont {Wiebe}}, \bibinfo
  {author} {\bibfnamefont {C.}~\bibnamefont {Gidney}}, \bibinfo {author}
  {\bibfnamefont {A.}~\bibnamefont {Aspuru-Guzik}}, \bibinfo {author}
  {\bibfnamefont {G.~K.-L.}\ \bibnamefont {Chan}}, \ and\ \bibinfo {author}
  {\bibfnamefont {R.}~\bibnamefont {Babbush}},\ }\bibfield  {title} {\emph
  {\bibinfo {title} {Quantum simulation of electronic structure with linear
  depth and connectivity},\ }}\href {\doibase 10.1103/PhysRevLett.120.110501}
  {\bibfield  {journal} {\bibinfo  {journal} {Phys. Rev. Lett.}\ }\textbf
  {\bibinfo {volume} {120}},\ \bibinfo {pages} {110501} (\bibinfo {year}
  {2018})}\BibitemShut {NoStop}%
\bibitem [{\citenamefont {Harper}\ \emph {et~al.}(2020)\citenamefont {Harper},
  \citenamefont {Flammia},\ and\ \citenamefont {Wallman}}]{QuantumNoise}%
  \BibitemOpen
  \bibfield  {author} {\bibinfo {author} {\bibfnamefont {R.}~\bibnamefont
  {Harper}}, \bibinfo {author} {\bibfnamefont {S.~T.}\ \bibnamefont {Flammia}},
  \ and\ \bibinfo {author} {\bibfnamefont {J.~J.}\ \bibnamefont {Wallman}},\
  }\bibfield  {title} {\emph {\bibinfo {title} {Efficient learning of quantum
  noise},\ }}\href {\doibase https://doi.org/10.1038/s41567-020-0992-8}
  {\bibfield  {journal} {\bibinfo  {journal} {Nat. Phys.}\ }\textbf {\bibinfo
  {volume} {16}},\ \bibinfo {pages} {1184} (\bibinfo {year}
  {2020})}\BibitemShut {NoStop}%
\bibitem [{\citenamefont {Song}\ \emph {et~al.}(2017)\citenamefont {Song},
  \citenamefont {Zheng}, \citenamefont {Zhang} \emph {et~al.}}]{QuantumNoise2}%
  \BibitemOpen
  \bibfield  {author} {\bibinfo {author} {\bibfnamefont {C.}~\bibnamefont
  {Song}}, \bibinfo {author} {\bibfnamefont {S.-B.}\ \bibnamefont {Zheng}},
  \bibinfo {author} {\bibfnamefont {P.}~\bibnamefont {Zhang}},  \emph
  {et~al.},\ }\bibfield  {title} {\emph {\bibinfo {title} {Continuous-variable
  geometric phase and its manipulation for quantum computation in a
  superconducting circuit},\ }}\href {\doibase
  https://doi.org/10.1038/s41467-017-01156-5} {\bibfield  {journal} {\bibinfo
  {journal} {Nat Commun}\ }\textbf {\bibinfo {volume} {8}},\ \bibinfo {pages}
  {1} (\bibinfo {year} {2017})}\BibitemShut {NoStop}%
\bibitem [{\citenamefont {Berry}(1984)}]{BerryPhase}%
  \BibitemOpen
  \bibfield  {author} {\bibinfo {author} {\bibfnamefont {M.~V.}\ \bibnamefont
  {Berry}},\ }\bibfield  {title} {\emph {\bibinfo {title} {Quantal phase
  factors accompanying adiabatic changes},\ }}\href {\doibase
  10.1098/rspa.1984.0023} {\bibfield  {journal} {\bibinfo  {journal} {Proc. R.
  Soc. Lond. A}\ }\textbf {\bibinfo {volume} {392}},\ \bibinfo {pages} {45}
  (\bibinfo {year} {1984})}\BibitemShut {NoStop}%
\bibitem [{\citenamefont {Aharonov}\ and\ \citenamefont
  {Anandan}(1987)}]{AAphase}%
  \BibitemOpen
  \bibfield  {author} {\bibinfo {author} {\bibfnamefont {Y.}~\bibnamefont
  {Aharonov}}\ and\ \bibinfo {author} {\bibfnamefont {J.}~\bibnamefont
  {Anandan}},\ }\bibfield  {title} {\emph {\bibinfo {title} {Phase change
  during a cyclic quantum evolution},\ }}\href {\doibase
  10.1103/PhysRevLett.58.1593} {\bibfield  {journal} {\bibinfo  {journal}
  {Phys. Rev. Lett.}\ }\textbf {\bibinfo {volume} {58}},\ \bibinfo {pages}
  {1593} (\bibinfo {year} {1987})}\BibitemShut {NoStop}%
\bibitem [{\citenamefont {Samuel}\ and\ \citenamefont
  {Bhandari}(1988)}]{GeometricPhase1}%
  \BibitemOpen
  \bibfield  {author} {\bibinfo {author} {\bibfnamefont {J.}~\bibnamefont
  {Samuel}}\ and\ \bibinfo {author} {\bibfnamefont {R.}~\bibnamefont
  {Bhandari}},\ }\bibfield  {title} {\emph {\bibinfo {title} {General setting
  for berry's phase},\ }}\href {\doibase 10.1103/PhysRevLett.60.2339}
  {\bibfield  {journal} {\bibinfo  {journal} {Phys. Rev. Lett.}\ }\textbf
  {\bibinfo {volume} {60}},\ \bibinfo {pages} {2339} (\bibinfo {year}
  {1988})}\BibitemShut {NoStop}%
\bibitem [{\citenamefont {Wilczek}\ and\ \citenamefont
  {Zee}(1984)}]{GeometricPhase2}%
  \BibitemOpen
  \bibfield  {author} {\bibinfo {author} {\bibfnamefont {F.}~\bibnamefont
  {Wilczek}}\ and\ \bibinfo {author} {\bibfnamefont {A.}~\bibnamefont {Zee}},\
  }\bibfield  {title} {\emph {\bibinfo {title} {Appearance of gauge structure
  in simple dynamical systems},\ }}\href {\doibase 10.1103/PhysRevLett.52.2111}
  {\bibfield  {journal} {\bibinfo  {journal} {Phys. Rev. Lett.}\ }\textbf
  {\bibinfo {volume} {52}},\ \bibinfo {pages} {2111} (\bibinfo {year}
  {1984})}\BibitemShut {NoStop}%
\bibitem [{\citenamefont {Anandan}(1988)}]{GeometricPhase3}%
  \BibitemOpen
  \bibfield  {author} {\bibinfo {author} {\bibfnamefont {J.}~\bibnamefont
  {Anandan}},\ }\bibfield  {title} {\emph {\bibinfo {title} {Non-adiabatic
  non-abelian geometric phase},\ }}\href {\doibase
  https://doi.org/10.1016/0375-9601(88)91010-9} {\bibfield  {journal} {\bibinfo
   {journal} {Phys. Lett. A}\ }\textbf {\bibinfo {volume} {133}},\ \bibinfo
  {pages} {171} (\bibinfo {year} {1988})}\BibitemShut {NoStop}%
\bibitem [{\citenamefont {Leibfried}\ \emph {et~al.}(2003)\citenamefont
  {Leibfried}, \citenamefont {DeMarco}, \citenamefont {Meyer}, \citenamefont
  {Lucas}, \citenamefont {Barrett}, \citenamefont {Britton}, \citenamefont
  {Itano}, \citenamefont {Jelenkovi{\'c}}, \citenamefont {Langer},
  \citenamefont {Rosenband} \emph {et~al.}}]{TrappedIon1}%
  \BibitemOpen
  \bibfield  {author} {\bibinfo {author} {\bibfnamefont {D.}~\bibnamefont
  {Leibfried}}, \bibinfo {author} {\bibfnamefont {B.}~\bibnamefont {DeMarco}},
  \bibinfo {author} {\bibfnamefont {V.}~\bibnamefont {Meyer}}, \bibinfo
  {author} {\bibfnamefont {D.}~\bibnamefont {Lucas}}, \bibinfo {author}
  {\bibfnamefont {M.}~\bibnamefont {Barrett}}, \bibinfo {author} {\bibfnamefont
  {J.}~\bibnamefont {Britton}}, \bibinfo {author} {\bibfnamefont {W.~M.}\
  \bibnamefont {Itano}}, \bibinfo {author} {\bibfnamefont {B.}~\bibnamefont
  {Jelenkovi{\'c}}}, \bibinfo {author} {\bibfnamefont {C.}~\bibnamefont
  {Langer}}, \bibinfo {author} {\bibfnamefont {T.}~\bibnamefont {Rosenband}},
  \emph {et~al.},\ }\bibfield  {title} {\emph {\bibinfo {title} {Experimental
  demonstration of a robust, high-fidelity geometric two ion-qubit phase
  gate},\ }}\href {\doibase https://doi.org/10.1038/nature01492} {\bibfield
  {journal} {\bibinfo  {journal} {Nature}\ }\textbf {\bibinfo {volume} {422}},\
  \bibinfo {pages} {412} (\bibinfo {year} {2003})}\BibitemShut {NoStop}%
\bibitem [{\citenamefont {Cui}\ \emph {et~al.}(2019)\citenamefont {Cui},
  \citenamefont {Ai}, \citenamefont {He}, \citenamefont {Qian}, \citenamefont
  {Qin}, \citenamefont {Huang}, \citenamefont {Zhou}, \citenamefont {Li},
  \citenamefont {Tu},\ and\ \citenamefont {Guo}}]{TrappedIon2}%
  \BibitemOpen
  \bibfield  {author} {\bibinfo {author} {\bibfnamefont {J.-M.}\ \bibnamefont
  {Cui}}, \bibinfo {author} {\bibfnamefont {M.-Z.}\ \bibnamefont {Ai}},
  \bibinfo {author} {\bibfnamefont {R.}~\bibnamefont {He}}, \bibinfo {author}
  {\bibfnamefont {Z.-H.}\ \bibnamefont {Qian}}, \bibinfo {author}
  {\bibfnamefont {X.-K.}\ \bibnamefont {Qin}}, \bibinfo {author} {\bibfnamefont
  {Y.-F.}\ \bibnamefont {Huang}}, \bibinfo {author} {\bibfnamefont {Z.-W.}\
  \bibnamefont {Zhou}}, \bibinfo {author} {\bibfnamefont {C.-F.}\ \bibnamefont
  {Li}}, \bibinfo {author} {\bibfnamefont {T.}~\bibnamefont {Tu}}, \ and\
  \bibinfo {author} {\bibfnamefont {G.-C.}\ \bibnamefont {Guo}},\ }\bibfield
  {title} {\emph {\bibinfo {title} {Experimental demonstration of suppressing
  residual geometric dephasing},\ }}\href {\doibase
  https://doi.org/10.1016/j.scib.2019.09.007} {\bibfield  {journal} {\bibinfo
  {journal} {Sci. Bull.}\ }\textbf {\bibinfo {volume} {64}},\ \bibinfo {pages}
  {1757} (\bibinfo {year} {2019})}\BibitemShut {NoStop}%
\bibitem [{\citenamefont {Jones}\ \emph {et~al.}(2000)\citenamefont {Jones},
  \citenamefont {Vedral}, \citenamefont {Ekert},\ and\ \citenamefont
  {Castagnoli}}]{NMR1}%
  \BibitemOpen
  \bibfield  {author} {\bibinfo {author} {\bibfnamefont {J.~A.}\ \bibnamefont
  {Jones}}, \bibinfo {author} {\bibfnamefont {V.}~\bibnamefont {Vedral}},
  \bibinfo {author} {\bibfnamefont {A.}~\bibnamefont {Ekert}}, \ and\ \bibinfo
  {author} {\bibfnamefont {G.}~\bibnamefont {Castagnoli}},\ }\bibfield  {title}
  {\emph {\bibinfo {title} {Geometric quantum computation using nuclear
  magnetic resonance},\ }}\href {\doibase https://doi.org/10.1038/35002528}
  {\bibfield  {journal} {\bibinfo  {journal} {Nature}\ }\textbf {\bibinfo
  {volume} {403}},\ \bibinfo {pages} {869} (\bibinfo {year}
  {2000})}\BibitemShut {NoStop}%
\bibitem [{\citenamefont {Feng}\ \emph {et~al.}(2013)\citenamefont {Feng},
  \citenamefont {Xu},\ and\ \citenamefont {Long}}]{NMR2}%
  \BibitemOpen
  \bibfield  {author} {\bibinfo {author} {\bibfnamefont {G.}~\bibnamefont
  {Feng}}, \bibinfo {author} {\bibfnamefont {G.}~\bibnamefont {Xu}}, \ and\
  \bibinfo {author} {\bibfnamefont {G.}~\bibnamefont {Long}},\ }\bibfield
  {title} {\emph {\bibinfo {title} {Experimental realization of nonadiabatic
  holonomic quantum computation},\ }}\href {\doibase
  10.1103/PhysRevLett.110.190501} {\bibfield  {journal} {\bibinfo  {journal}
  {Phys. Rev. Lett.}\ }\textbf {\bibinfo {volume} {110}},\ \bibinfo {pages}
  {190501} (\bibinfo {year} {2013})}\BibitemShut {NoStop}%
\bibitem [{\citenamefont {Abdumalikov~Jr}\ \emph {et~al.}(2013)\citenamefont
  {Abdumalikov~Jr}, \citenamefont {Fink}, \citenamefont {Juliusson},
  \citenamefont {Pechal}, \citenamefont {Berger}, \citenamefont {Wallraff},\
  and\ \citenamefont {Filipp}}]{SuperconductingCircuits1}%
  \BibitemOpen
  \bibfield  {author} {\bibinfo {author} {\bibfnamefont {A.~A.}\ \bibnamefont
  {Abdumalikov~Jr}}, \bibinfo {author} {\bibfnamefont {J.~M.}\ \bibnamefont
  {Fink}}, \bibinfo {author} {\bibfnamefont {K.}~\bibnamefont {Juliusson}},
  \bibinfo {author} {\bibfnamefont {M.}~\bibnamefont {Pechal}}, \bibinfo
  {author} {\bibfnamefont {S.}~\bibnamefont {Berger}}, \bibinfo {author}
  {\bibfnamefont {A.}~\bibnamefont {Wallraff}}, \ and\ \bibinfo {author}
  {\bibfnamefont {S.}~\bibnamefont {Filipp}},\ }\bibfield  {title} {\emph
  {\bibinfo {title} {Experimental realization of non-abelian non-adiabatic
  geometric gates},\ }}\href {\doibase https://doi.org/10.1038/nature12010}
  {\bibfield  {journal} {\bibinfo  {journal} {Nature}\ }\textbf {\bibinfo
  {volume} {496}},\ \bibinfo {pages} {482} (\bibinfo {year}
  {2013})}\BibitemShut {NoStop}%
\bibitem [{\citenamefont {Xu}\ \emph {et~al.}(2018)\citenamefont {Xu},
  \citenamefont {Cai}, \citenamefont {Ma}, \citenamefont {Mu}, \citenamefont
  {Hu}, \citenamefont {Chen}, \citenamefont {Wang}, \citenamefont {Song},
  \citenamefont {Xue}, \citenamefont {Yin},\ and\ \citenamefont
  {Sun}}]{SuperconductingCircuits2}%
  \BibitemOpen
  \bibfield  {author} {\bibinfo {author} {\bibfnamefont {Y.}~\bibnamefont
  {Xu}}, \bibinfo {author} {\bibfnamefont {W.}~\bibnamefont {Cai}}, \bibinfo
  {author} {\bibfnamefont {Y.}~\bibnamefont {Ma}}, \bibinfo {author}
  {\bibfnamefont {X.}~\bibnamefont {Mu}}, \bibinfo {author} {\bibfnamefont
  {L.}~\bibnamefont {Hu}}, \bibinfo {author} {\bibfnamefont {T.}~\bibnamefont
  {Chen}}, \bibinfo {author} {\bibfnamefont {H.}~\bibnamefont {Wang}}, \bibinfo
  {author} {\bibfnamefont {Y.~P.}\ \bibnamefont {Song}}, \bibinfo {author}
  {\bibfnamefont {Z.-Y.}\ \bibnamefont {Xue}}, \bibinfo {author} {\bibfnamefont
  {Z.-q.}\ \bibnamefont {Yin}}, \ and\ \bibinfo {author} {\bibfnamefont
  {L.}~\bibnamefont {Sun}},\ }\bibfield  {title} {\emph {\bibinfo {title}
  {Single-loop realization of arbitrary nonadiabatic holonomic single-qubit
  quantum gates in a superconducting circuit},\ }}\href {\doibase
  10.1103/PhysRevLett.121.110501} {\bibfield  {journal} {\bibinfo  {journal}
  {Phys. Rev. Lett.}\ }\textbf {\bibinfo {volume} {121}},\ \bibinfo {pages}
  {110501} (\bibinfo {year} {2018})}\BibitemShut {NoStop}%
\bibitem [{\citenamefont {Zhou}\ \emph {et~al.}(2021)\citenamefont {Zhou},
  \citenamefont {Li}, \citenamefont {Pan}, \citenamefont {Zhang}, \citenamefont
  {Chen},\ and\ \citenamefont {Xue}}]{RobustNGQC2}%
  \BibitemOpen
  \bibfield  {author} {\bibinfo {author} {\bibfnamefont {J.}~\bibnamefont
  {Zhou}}, \bibinfo {author} {\bibfnamefont {S.}~\bibnamefont {Li}}, \bibinfo
  {author} {\bibfnamefont {G.-Z.}\ \bibnamefont {Pan}}, \bibinfo {author}
  {\bibfnamefont {G.}~\bibnamefont {Zhang}}, \bibinfo {author} {\bibfnamefont
  {T.}~\bibnamefont {Chen}}, \ and\ \bibinfo {author} {\bibfnamefont {Z.-Y.}\
  \bibnamefont {Xue}},\ }\bibfield  {title} {\emph {\bibinfo {title}
  {Nonadiabatic geometric quantum gates that are insensitive to qubit-frequency
  drifts},\ }}\href {\doibase 10.1103/PhysRevA.103.032609} {\bibfield
  {journal} {\bibinfo  {journal} {Phys. Rev. A}\ }\textbf {\bibinfo {volume}
  {103}},\ \bibinfo {pages} {032609} (\bibinfo {year} {2021})}\BibitemShut
  {NoStop}%
\bibitem [{\citenamefont {Liang}\ and\ \citenamefont
  {Xue}(2022)}]{RobustNGQC1}%
  \BibitemOpen
  \bibfield  {author} {\bibinfo {author} {\bibfnamefont {M.-J.}\ \bibnamefont
  {Liang}}\ and\ \bibinfo {author} {\bibfnamefont {Z.-Y.}\ \bibnamefont
  {Xue}},\ }\bibfield  {title} {\emph {\bibinfo {title} {Robust nonadiabatic
  geometric quantum computation by dynamical correction},\ }}\href {\doibase
  10.1103/PhysRevA.106.012603} {\bibfield  {journal} {\bibinfo  {journal}
  {Phys. Rev. A}\ }\textbf {\bibinfo {volume} {106}},\ \bibinfo {pages}
  {012603} (\bibinfo {year} {2022})}\BibitemShut {NoStop}%
\bibitem [{\citenamefont {Wang}\ and\ \citenamefont
  {Keiji}(2001)}]{AAphaseQuantumGate}%
  \BibitemOpen
  \bibfield  {author} {\bibinfo {author} {\bibfnamefont {X.~B.}\ \bibnamefont
  {Wang}}\ and\ \bibinfo {author} {\bibfnamefont {M.}~\bibnamefont {Keiji}},\
  }\bibfield  {title} {\emph {\bibinfo {title} {Nonadiabatic conditional
  geometric phase shift with nmr},\ }}\href {\doibase
  10.1103/PhysRevLett.87.097901} {\bibfield  {journal} {\bibinfo  {journal}
  {Phys. Rev. Lett.}\ }\textbf {\bibinfo {volume} {87}},\ \bibinfo {pages}
  {097901} (\bibinfo {year} {2001})}\BibitemShut {NoStop}%
\bibitem [{\citenamefont {Berry}(2009)}]{TQD}%
  \BibitemOpen
  \bibfield  {author} {\bibinfo {author} {\bibfnamefont {M.~V.}\ \bibnamefont
  {Berry}},\ }\bibfield  {title} {\emph {\bibinfo {title} {Transitionless
  quantum driving},\ }}\href {\doibase 10.1088/1751-8113/42/36/365303}
  {\bibfield  {journal} {\bibinfo  {journal} {J. Phys. A: Math. Theor.}\
  }\textbf {\bibinfo {volume} {42}},\ \bibinfo {pages} {365303} (\bibinfo
  {year} {2009})}\BibitemShut {NoStop}%
\bibitem [{\citenamefont {Viola}\ and\ \citenamefont {Lloyd}(1998)}]{DD1}%
  \BibitemOpen
  \bibfield  {author} {\bibinfo {author} {\bibfnamefont {L.}~\bibnamefont
  {Viola}}\ and\ \bibinfo {author} {\bibfnamefont {S.}~\bibnamefont {Lloyd}},\
  }\bibfield  {title} {\emph {\bibinfo {title} {Dynamical suppression of
  decoherence in two-state quantum systems},\ }}\href {\doibase
  10.1103/PhysRevA.58.2733} {\bibfield  {journal} {\bibinfo  {journal} {Phys.
  Rev. A}\ }\textbf {\bibinfo {volume} {58}},\ \bibinfo {pages} {2733}
  (\bibinfo {year} {1998})}\BibitemShut {NoStop}%
\bibitem [{\citenamefont {Witzel}\ and\ \citenamefont
  {Sarma}(2007)}]{MultipleCPMG}%
  \BibitemOpen
  \bibfield  {author} {\bibinfo {author} {\bibfnamefont {W.~M.}\ \bibnamefont
  {Witzel}}\ and\ \bibinfo {author} {\bibfnamefont {S.~D.}\ \bibnamefont
  {Sarma}},\ }\bibfield  {title} {\emph {\bibinfo {title} {Multiple-pulse
  coherence enhancement of solid state spin qubits},\ }}\href {\doibase
  10.1103/PhysRevLett.98.077601} {\bibfield  {journal} {\bibinfo  {journal}
  {Phys. Rev. Lett.}\ }\textbf {\bibinfo {volume} {98}},\ \bibinfo {pages}
  {077601} (\bibinfo {year} {2007})}\BibitemShut {NoStop}%
\bibitem [{\citenamefont {Uhrig}(2007)}]{UDD1}%
  \BibitemOpen
  \bibfield  {author} {\bibinfo {author} {\bibfnamefont {G.~S.}\ \bibnamefont
  {Uhrig}},\ }\bibfield  {title} {\emph {\bibinfo {title} {Keeping a quantum
  bit alive by optimized $\ensuremath{\pi}$-pulse sequences},\ }}\href
  {\doibase 10.1103/PhysRevLett.98.100504} {\bibfield  {journal} {\bibinfo
  {journal} {Phys. Rev. Lett.}\ }\textbf {\bibinfo {volume} {98}},\ \bibinfo
  {pages} {100504} (\bibinfo {year} {2007})}\BibitemShut {NoStop}%
\bibitem [{\citenamefont {Pasini}\ and\ \citenamefont {Uhrig}(2010)}]{UDD2}%
  \BibitemOpen
  \bibfield  {author} {\bibinfo {author} {\bibfnamefont {S.}~\bibnamefont
  {Pasini}}\ and\ \bibinfo {author} {\bibfnamefont {G.~S.}\ \bibnamefont
  {Uhrig}},\ }\bibfield  {title} {\emph {\bibinfo {title} {Optimized dynamical
  decoupling for power-law noise spectra},\ }}\href {\doibase
  10.1103/PhysRevA.81.012309} {\bibfield  {journal} {\bibinfo  {journal} {Phys.
  Rev. A}\ }\textbf {\bibinfo {volume} {81}},\ \bibinfo {pages} {012309}
  (\bibinfo {year} {2010})}\BibitemShut {NoStop}%
\bibitem [{\citenamefont {Qin}\ \emph {et~al.}(2017)\citenamefont {Qin},
  \citenamefont {Guo},\ and\ \citenamefont {Zhou}}]{GaussianNoiseForm2}%
  \BibitemOpen
  \bibfield  {author} {\bibinfo {author} {\bibfnamefont {X.-K.}\ \bibnamefont
  {Qin}}, \bibinfo {author} {\bibfnamefont {G.-C.}\ \bibnamefont {Guo}}, \ and\
  \bibinfo {author} {\bibfnamefont {Z.-W.}\ \bibnamefont {Zhou}},\ }\bibfield
  {title} {\emph {\bibinfo {title} {Suppressing the geometric dephasing of
  berry phase by using modified dynamical decoupling sequences},\ }}\href
  {\doibase 10.1088/1367-2630/aa5488} {\bibfield  {journal} {\bibinfo
  {journal} {New J. Phys.}\ }\textbf {\bibinfo {volume} {19}},\ \bibinfo
  {pages} {013025} (\bibinfo {year} {2017})}\BibitemShut {NoStop}%
\bibitem [{\citenamefont {Zhu}\ and\ \citenamefont
  {Wang}(2003)}]{UnconventionalGQC1}%
  \BibitemOpen
  \bibfield  {author} {\bibinfo {author} {\bibfnamefont {S.-L.}\ \bibnamefont
  {Zhu}}\ and\ \bibinfo {author} {\bibfnamefont {Z.~D.}\ \bibnamefont {Wang}},\
  }\bibfield  {title} {\emph {\bibinfo {title} {Unconventional geometric
  quantum computation},\ }}\href {\doibase 10.1103/PhysRevLett.91.187902}
  {\bibfield  {journal} {\bibinfo  {journal} {Phys. Rev. Lett.}\ }\textbf
  {\bibinfo {volume} {91}},\ \bibinfo {pages} {187902} (\bibinfo {year}
  {2003})}\BibitemShut {NoStop}%
\bibitem [{\citenamefont {Du}\ \emph {et~al.}(2006)\citenamefont {Du},
  \citenamefont {Zou},\ and\ \citenamefont {Wang}}]{UnconventionalGQC2}%
  \BibitemOpen
  \bibfield  {author} {\bibinfo {author} {\bibfnamefont {J.}~\bibnamefont
  {Du}}, \bibinfo {author} {\bibfnamefont {P.}~\bibnamefont {Zou}}, \ and\
  \bibinfo {author} {\bibfnamefont {Z.~D.}\ \bibnamefont {Wang}},\ }\bibfield
  {title} {\emph {\bibinfo {title} {Experimental implementation of
  high-fidelity unconventional geometric quantum gates using an nmr
  interferometer},\ }}\href {\doibase 10.1103/PhysRevA.74.020302} {\bibfield
  {journal} {\bibinfo  {journal} {Phys. Rev. A}\ }\textbf {\bibinfo {volume}
  {74}},\ \bibinfo {pages} {020302} (\bibinfo {year} {2006})}\BibitemShut
  {NoStop}%
\bibitem [{\citenamefont {Wang}\ \emph {et~al.}(2007)\citenamefont {Wang},
  \citenamefont {Wu}, \citenamefont {Feng}, \citenamefont {Kwek}, \citenamefont
  {Lai}, \citenamefont {Oh},\ and\ \citenamefont
  {Vedral}}]{UnconventionalGQC3}%
  \BibitemOpen
  \bibfield  {author} {\bibinfo {author} {\bibfnamefont {Z.~S.}\ \bibnamefont
  {Wang}}, \bibinfo {author} {\bibfnamefont {C.}~\bibnamefont {Wu}}, \bibinfo
  {author} {\bibfnamefont {X.-L.}\ \bibnamefont {Feng}}, \bibinfo {author}
  {\bibfnamefont {L.~C.}\ \bibnamefont {Kwek}}, \bibinfo {author}
  {\bibfnamefont {C.~H.}\ \bibnamefont {Lai}}, \bibinfo {author} {\bibfnamefont
  {C.~H.}\ \bibnamefont {Oh}}, \ and\ \bibinfo {author} {\bibfnamefont
  {V.}~\bibnamefont {Vedral}},\ }\bibfield  {title} {\emph {\bibinfo {title}
  {Nonadiabatic geometric quantum computation},\ }}\href {\doibase
  10.1103/PhysRevA.76.044303} {\bibfield  {journal} {\bibinfo  {journal} {Phys.
  Rev. A}\ }\textbf {\bibinfo {volume} {76}},\ \bibinfo {pages} {044303}
  (\bibinfo {year} {2007})}\BibitemShut {NoStop}%
\bibitem [{\citenamefont {Feng}\ \emph {et~al.}(2009)\citenamefont {Feng},
  \citenamefont {Wu}, \citenamefont {Sun},\ and\ \citenamefont
  {Oh}}]{DecoherenceFreeSubspace}%
  \BibitemOpen
  \bibfield  {author} {\bibinfo {author} {\bibfnamefont {X.-L.}\ \bibnamefont
  {Feng}}, \bibinfo {author} {\bibfnamefont {C.}~\bibnamefont {Wu}}, \bibinfo
  {author} {\bibfnamefont {H.}~\bibnamefont {Sun}}, \ and\ \bibinfo {author}
  {\bibfnamefont {C.~H.}\ \bibnamefont {Oh}},\ }\bibfield  {title} {\emph
  {\bibinfo {title} {Geometric entangling gates in decoherence-free subspaces
  with minimal requirements},\ }}\href {\doibase
  10.1103/PhysRevLett.103.200501} {\bibfield  {journal} {\bibinfo  {journal}
  {Phys. Rev. Lett.}\ }\textbf {\bibinfo {volume} {103}},\ \bibinfo {pages}
  {200501} (\bibinfo {year} {2009})}\BibitemShut {NoStop}%
\bibitem [{\citenamefont {Oreshkov}\ \emph
  {et~al.}(2009{\natexlab{a}})\citenamefont {Oreshkov}, \citenamefont {Brun},\
  and\ \citenamefont {Lidar}}]{QuantumErrorCorrection1}%
  \BibitemOpen
  \bibfield  {author} {\bibinfo {author} {\bibfnamefont {O.}~\bibnamefont
  {Oreshkov}}, \bibinfo {author} {\bibfnamefont {T.~A.}\ \bibnamefont {Brun}},
  \ and\ \bibinfo {author} {\bibfnamefont {D.~A.}\ \bibnamefont {Lidar}},\
  }\bibfield  {title} {\emph {\bibinfo {title} {Fault-tolerant holonomic
  quantum computation},\ }}\href {\doibase 10.1103/PhysRevLett.102.070502}
  {\bibfield  {journal} {\bibinfo  {journal} {Phys. Rev. Lett.}\ }\textbf
  {\bibinfo {volume} {102}},\ \bibinfo {pages} {070502} (\bibinfo {year}
  {2009}{\natexlab{a}})}\BibitemShut {NoStop}%
\bibitem [{\citenamefont {Oreshkov}\ \emph
  {et~al.}(2009{\natexlab{b}})\citenamefont {Oreshkov}, \citenamefont {Brun},\
  and\ \citenamefont {Lidar}}]{QuantumErrorCorrection2}%
  \BibitemOpen
  \bibfield  {author} {\bibinfo {author} {\bibfnamefont {O.}~\bibnamefont
  {Oreshkov}}, \bibinfo {author} {\bibfnamefont {T.~A.}\ \bibnamefont {Brun}},
  \ and\ \bibinfo {author} {\bibfnamefont {D.~A.}\ \bibnamefont {Lidar}},\
  }\bibfield  {title} {\emph {\bibinfo {title} {Scheme for fault-tolerant
  holonomic computation on stabilizer codes},\ }}\href {\doibase
  10.1103/PhysRevA.80.022325} {\bibfield  {journal} {\bibinfo  {journal} {Phys.
  Rev. A}\ }\textbf {\bibinfo {volume} {80}},\ \bibinfo {pages} {022325}
  (\bibinfo {year} {2009}{\natexlab{b}})}\BibitemShut {NoStop}%
\bibitem [{\citenamefont {Gardiner}\ \emph {et~al.}(2004)\citenamefont
  {Gardiner}, \citenamefont {Zoller},\ and\ \citenamefont
  {Zoller}}]{StochasticNoise}%
  \BibitemOpen
  \bibfield  {author} {\bibinfo {author} {\bibfnamefont {C.}~\bibnamefont
  {Gardiner}}, \bibinfo {author} {\bibfnamefont {P.}~\bibnamefont {Zoller}}, \
  and\ \bibinfo {author} {\bibfnamefont {P.}~\bibnamefont {Zoller}},\
  }\href@noop {} {\emph {\bibinfo {title} {Quantum noise: a handbook of
  Markovian and non-Markovian quantum stochastic methods with applications to
  quantum optics}}}\ (\bibinfo  {publisher} {Springer Science \& Business
  Media},\ \bibinfo {year} {2004})\BibitemShut {NoStop}%
\bibitem [{\citenamefont {Gerry}\ \emph {et~al.}(2005)\citenamefont {Gerry},
  \citenamefont {Knight},\ and\ \citenamefont {Knight}}]{RabiModel}%
  \BibitemOpen
  \bibfield  {author} {\bibinfo {author} {\bibfnamefont {C.}~\bibnamefont
  {Gerry}}, \bibinfo {author} {\bibfnamefont {P.}~\bibnamefont {Knight}}, \
  and\ \bibinfo {author} {\bibfnamefont {P.~L.}\ \bibnamefont {Knight}},\
  }\href@noop {} {\emph {\bibinfo {title} {Introductory quantum optics}}}\
  (\bibinfo  {publisher} {Cambridge university press},\ \bibinfo {year}
  {2005})\BibitemShut {NoStop}%
\bibitem [{\citenamefont {Leek}\ \emph {et~al.}(2007)\citenamefont {Leek},
  \citenamefont {Fink}, \citenamefont {Blais}, \citenamefont {Bianchetti},
  \citenamefont {Göppl}, \citenamefont {Gambetta}, \citenamefont {Schuster},
  \citenamefont {Frunzio}, \citenamefont {Schoelkopf},\ and\ \citenamefont
  {Wallraff}}]{RWA}%
  \BibitemOpen
  \bibfield  {author} {\bibinfo {author} {\bibfnamefont {P.~J.}\ \bibnamefont
  {Leek}}, \bibinfo {author} {\bibfnamefont {J.~M.}\ \bibnamefont {Fink}},
  \bibinfo {author} {\bibfnamefont {A.}~\bibnamefont {Blais}}, \bibinfo
  {author} {\bibfnamefont {R.}~\bibnamefont {Bianchetti}}, \bibinfo {author}
  {\bibfnamefont {M.}~\bibnamefont {Göppl}}, \bibinfo {author} {\bibfnamefont
  {J.~M.}\ \bibnamefont {Gambetta}}, \bibinfo {author} {\bibfnamefont {D.~I.}\
  \bibnamefont {Schuster}}, \bibinfo {author} {\bibfnamefont {L.}~\bibnamefont
  {Frunzio}}, \bibinfo {author} {\bibfnamefont {R.~J.}\ \bibnamefont
  {Schoelkopf}}, \ and\ \bibinfo {author} {\bibfnamefont {A.}~\bibnamefont
  {Wallraff}},\ }\bibfield  {title} {\emph {\bibinfo {title} {Observation of
  berry's phase in a solid-state qubit},\ }}\href {\doibase
  10.1126/science.1149858} {\bibfield  {journal} {\bibinfo  {journal}
  {Science}\ }\textbf {\bibinfo {volume} {318}},\ \bibinfo {pages} {1889}
  (\bibinfo {year} {2007})}\BibitemShut {NoStop}%
\bibitem [{\citenamefont {Zheng}\ \emph {et~al.}(2008)\citenamefont {Zheng},
  \citenamefont {Zhu},\ and\ \citenamefont {Zubairy}}]{StrongDrivingDynamics1}%
  \BibitemOpen
  \bibfield  {author} {\bibinfo {author} {\bibfnamefont {H.}~\bibnamefont
  {Zheng}}, \bibinfo {author} {\bibfnamefont {S.~Y.}\ \bibnamefont {Zhu}}, \
  and\ \bibinfo {author} {\bibfnamefont {M.~S.}\ \bibnamefont {Zubairy}},\
  }\bibfield  {title} {\emph {\bibinfo {title} {Quantum zeno and anti-zeno
  effects: Without the rotating-wave approximation},\ }}\href {\doibase
  10.1103/PhysRevLett.101.200404} {\bibfield  {journal} {\bibinfo  {journal}
  {Phys. Rev. Lett.}\ }\textbf {\bibinfo {volume} {101}},\ \bibinfo {pages}
  {200404} (\bibinfo {year} {2008})}\BibitemShut {NoStop}%
\bibitem [{\citenamefont {Deng}\ \emph {et~al.}(2016)\citenamefont {Deng},
  \citenamefont {Shen}, \citenamefont {Ashhab},\ and\ \citenamefont
  {Lupascu}}]{StrongDrivingDynamics2}%
  \BibitemOpen
  \bibfield  {author} {\bibinfo {author} {\bibfnamefont {C.}~\bibnamefont
  {Deng}}, \bibinfo {author} {\bibfnamefont {F.}~\bibnamefont {Shen}}, \bibinfo
  {author} {\bibfnamefont {S.}~\bibnamefont {Ashhab}}, \ and\ \bibinfo {author}
  {\bibfnamefont {A.}~\bibnamefont {Lupascu}},\ }\bibfield  {title} {\emph
  {\bibinfo {title} {Dynamics of a two-level system under strong driving:
  Quantum-gate optimization based on floquet theory},\ }}\href {\doibase
  10.1103/PhysRevA.94.032323} {\bibfield  {journal} {\bibinfo  {journal} {Phys.
  Rev. A}\ }\textbf {\bibinfo {volume} {94}},\ \bibinfo {pages} {032323}
  (\bibinfo {year} {2016})}\BibitemShut {NoStop}%
\bibitem [{\citenamefont {L\"u}\ and\ \citenamefont
  {Zheng}(2012)}]{StrongDrivingDynamics3}%
  \BibitemOpen
  \bibfield  {author} {\bibinfo {author} {\bibfnamefont {Z.}~\bibnamefont
  {L\"u}}\ and\ \bibinfo {author} {\bibfnamefont {H.}~\bibnamefont {Zheng}},\
  }\bibfield  {title} {\emph {\bibinfo {title} {Effects of counter-rotating
  interaction on driven tunneling dynamics: Coherent destruction of tunneling
  and bloch-siegert shift},\ }}\href {\doibase 10.1103/PhysRevA.86.023831}
  {\bibfield  {journal} {\bibinfo  {journal} {Phys. Rev. A}\ }\textbf {\bibinfo
  {volume} {86}},\ \bibinfo {pages} {023831} (\bibinfo {year}
  {2012})}\BibitemShut {NoStop}%
\bibitem [{\citenamefont {Jing}\ \emph {et~al.}(2009)\citenamefont {Jing},
  \citenamefont {L\"u},\ and\ \citenamefont {Ficek}}]{StrongDrivingDynamics4}%
  \BibitemOpen
  \bibfield  {author} {\bibinfo {author} {\bibfnamefont {J.}~\bibnamefont
  {Jing}}, \bibinfo {author} {\bibfnamefont {Z.-G.}\ \bibnamefont {L\"u}}, \
  and\ \bibinfo {author} {\bibfnamefont {Z.}~\bibnamefont {Ficek}},\ }\bibfield
   {title} {\emph {\bibinfo {title} {Breakdown of the rotating-wave
  approximation in the description of entanglement of spin-anticorrelated
  states},\ }}\href {\doibase 10.1103/PhysRevA.79.044305} {\bibfield  {journal}
  {\bibinfo  {journal} {Phys. Rev. A}\ }\textbf {\bibinfo {volume} {79}},\
  \bibinfo {pages} {044305} (\bibinfo {year} {2009})}\BibitemShut {NoStop}%
\bibitem [{\citenamefont {Dong}\ \emph {et~al.}(2021)\citenamefont {Dong},
  \citenamefont {Zhuang}, \citenamefont {Economou},\ and\ \citenamefont
  {Barnes}}]{Solidangle}%
  \BibitemOpen
  \bibfield  {author} {\bibinfo {author} {\bibfnamefont {W.}~\bibnamefont
  {Dong}}, \bibinfo {author} {\bibfnamefont {F.}~\bibnamefont {Zhuang}},
  \bibinfo {author} {\bibfnamefont {S.~E.}\ \bibnamefont {Economou}}, \ and\
  \bibinfo {author} {\bibfnamefont {E.}~\bibnamefont {Barnes}},\ }\bibfield
  {title} {\emph {\bibinfo {title} {Doubly geometric quantum control},\ }}\href
  {\doibase 10.1103/PRXQuantum.2.030333} {\bibfield  {journal} {\bibinfo
  {journal} {PRX Quantum}\ }\textbf {\bibinfo {volume} {2}},\ \bibinfo {pages}
  {030333} (\bibinfo {year} {2021})}\BibitemShut {NoStop}%
\bibitem [{\citenamefont {Gregefalk}\ and\ \citenamefont
  {Sj\"oqvist}(2022)}]{AAphaseQuantumGate2}%
  \BibitemOpen
  \bibfield  {author} {\bibinfo {author} {\bibfnamefont {A.}~\bibnamefont
  {Gregefalk}}\ and\ \bibinfo {author} {\bibfnamefont {E.}~\bibnamefont
  {Sj\"oqvist}},\ }\bibfield  {title} {\emph {\bibinfo {title} {Transitionless
  quantum driving in spin echo},\ }}\href {\doibase
  10.1103/PhysRevApplied.17.024012} {\bibfield  {journal} {\bibinfo  {journal}
  {Phys. Rev. Applied}\ }\textbf {\bibinfo {volume} {17}},\ \bibinfo {pages}
  {024012} (\bibinfo {year} {2022})}\BibitemShut {NoStop}%
\bibitem [{\citenamefont {Cywi\ifmmode~\acute{n}\else \'{n}\fi{}ski}\ \emph
  {et~al.}(2008)\citenamefont {Cywi\ifmmode~\acute{n}\else \'{n}\fi{}ski},
  \citenamefont {Lutchyn}, \citenamefont {Nave},\ and\ \citenamefont
  {Das~Sarma}}]{PulseSequenceAnalysis2}%
  \BibitemOpen
  \bibfield  {author} {\bibinfo {author} {\bibfnamefont {L.}~\bibnamefont
  {Cywi\ifmmode~\acute{n}\else \'{n}\fi{}ski}}, \bibinfo {author}
  {\bibfnamefont {R.~M.}\ \bibnamefont {Lutchyn}}, \bibinfo {author}
  {\bibfnamefont {C.~P.}\ \bibnamefont {Nave}}, \ and\ \bibinfo {author}
  {\bibfnamefont {S.}~\bibnamefont {Das~Sarma}},\ }\bibfield  {title} {\emph
  {\bibinfo {title} {How to enhance dephasing time in superconducting qubits},\
  }}\href {\doibase 10.1103/PhysRevB.77.174509} {\bibfield  {journal} {\bibinfo
   {journal} {Phys. Rev. B}\ }\textbf {\bibinfo {volume} {77}},\ \bibinfo
  {pages} {174509} (\bibinfo {year} {2008})}\BibitemShut {NoStop}%
\bibitem [{\citenamefont {Sza{\'{n}}kowski}\ \emph {et~al.}(2017)\citenamefont
  {Sza{\'{n}}kowski}, \citenamefont {Ramon}, \citenamefont {Krzywda},
  \citenamefont {Kwiatkowski},\ and\ \citenamefont
  {Cywi{\'{n}}ski}}]{PulseSequenceAnalysis1}%
  \BibitemOpen
  \bibfield  {author} {\bibinfo {author} {\bibfnamefont {P.}~\bibnamefont
  {Sza{\'{n}}kowski}}, \bibinfo {author} {\bibfnamefont {G.}~\bibnamefont
  {Ramon}}, \bibinfo {author} {\bibfnamefont {J.}~\bibnamefont {Krzywda}},
  \bibinfo {author} {\bibfnamefont {D.}~\bibnamefont {Kwiatkowski}}, \ and\
  \bibinfo {author} {\bibfnamefont {{\L}.}~\bibnamefont {Cywi{\'{n}}ski}},\
  }\bibfield  {title} {\emph {\bibinfo {title} {Environmental noise
  spectroscopy with qubits subjected to dynamical decoupling},\ }}\href
  {\doibase 10.1088/1361-648x/aa7648} {\bibfield  {journal} {\bibinfo
  {journal} {J. Phys.: Condens. Matter}\ }\textbf {\bibinfo {volume} {29}},\
  \bibinfo {pages} {333001} (\bibinfo {year} {2017})}\BibitemShut {NoStop}%
\bibitem [{\citenamefont {Jing}\ \emph {et~al.}(2017)\citenamefont {Jing},
  \citenamefont {Lam},\ and\ \citenamefont {Wu}}]{GaussianNoiseForm1}%
  \BibitemOpen
  \bibfield  {author} {\bibinfo {author} {\bibfnamefont {J.}~\bibnamefont
  {Jing}}, \bibinfo {author} {\bibfnamefont {C.-H.}\ \bibnamefont {Lam}}, \
  and\ \bibinfo {author} {\bibfnamefont {L.-A.}\ \bibnamefont {Wu}},\
  }\bibfield  {title} {\emph {\bibinfo {title} {Non-abelian holonomic
  transformation in the presence of classical noise},\ }}\href {\doibase
  10.1103/PhysRevA.95.012334} {\bibfield  {journal} {\bibinfo  {journal} {Phys.
  Rev. A}\ }\textbf {\bibinfo {volume} {95}},\ \bibinfo {pages} {012334}
  (\bibinfo {year} {2017})}\BibitemShut {NoStop}%
\bibitem [{\citenamefont {Yang}\ \emph {et~al.}(2011)\citenamefont {Yang},
  \citenamefont {Wang},\ and\ \citenamefont {Liu}}]{ReviewsofDD}%
  \BibitemOpen
  \bibfield  {author} {\bibinfo {author} {\bibfnamefont {W.}~\bibnamefont
  {Yang}}, \bibinfo {author} {\bibfnamefont {Z.-Y.}\ \bibnamefont {Wang}}, \
  and\ \bibinfo {author} {\bibfnamefont {R.-B.}\ \bibnamefont {Liu}},\
  }\bibfield  {title} {\emph {\bibinfo {title} {Preserving qubit coherence by
  dynamical decoupling},\ }}\href {\doibase
  https://doi.org/10.1007/s11467-010-0113-8} {\bibfield  {journal} {\bibinfo
  {journal} {Front. Phys. China}\ }\textbf {\bibinfo {volume} {6}},\ \bibinfo
  {pages} {2} (\bibinfo {year} {2011})}\BibitemShut {NoStop}%
\bibitem [{\citenamefont {Hahn}(1950)}]{SpinEcho}%
  \BibitemOpen
  \bibfield  {author} {\bibinfo {author} {\bibfnamefont {E.~L.}\ \bibnamefont
  {Hahn}},\ }\bibfield  {title} {\emph {\bibinfo {title} {Spin echoes},\
  }}\href {\doibase 10.1103/PhysRev.80.580} {\bibfield  {journal} {\bibinfo
  {journal} {Phys. Rev.}\ }\textbf {\bibinfo {volume} {80}},\ \bibinfo {pages}
  {580} (\bibinfo {year} {1950})}\BibitemShut {NoStop}%
\bibitem [{\citenamefont {Carr}\ and\ \citenamefont {Purcell}(1954)}]{CPMG1}%
  \BibitemOpen
  \bibfield  {author} {\bibinfo {author} {\bibfnamefont {H.~Y.}\ \bibnamefont
  {Carr}}\ and\ \bibinfo {author} {\bibfnamefont {E.~M.}\ \bibnamefont
  {Purcell}},\ }\bibfield  {title} {\emph {\bibinfo {title} {Effects of
  diffusion on free precession in nuclear magnetic resonance experiments},\
  }}\href {\doibase 10.1103/PhysRev.94.630} {\bibfield  {journal} {\bibinfo
  {journal} {Phys. Rev.}\ }\textbf {\bibinfo {volume} {94}},\ \bibinfo {pages}
  {630} (\bibinfo {year} {1954})}\BibitemShut {NoStop}%
\bibitem [{\citenamefont {Meiboom}\ and\ \citenamefont {Gill}(1958)}]{CPMG2}%
  \BibitemOpen
  \bibfield  {author} {\bibinfo {author} {\bibfnamefont {S.}~\bibnamefont
  {Meiboom}}\ and\ \bibinfo {author} {\bibfnamefont {D.}~\bibnamefont {Gill}},\
  }\bibfield  {title} {\emph {\bibinfo {title} {Modified spin‐echo method for
  measuring nuclear relaxation times},\ }}\href {\doibase 10.1063/1.1716296}
  {\bibfield  {journal} {\bibinfo  {journal} {Rev. Sci. Instrum.}\ }\textbf
  {\bibinfo {volume} {29}},\ \bibinfo {pages} {688} (\bibinfo {year}
  {1958})}\BibitemShut {NoStop}%
\bibitem [{\citenamefont {Hatomura}\ and\ \citenamefont
  {Takahashi}(2021)}]{approTQD1}%
  \BibitemOpen
  \bibfield  {author} {\bibinfo {author} {\bibfnamefont {T.}~\bibnamefont
  {Hatomura}}\ and\ \bibinfo {author} {\bibfnamefont {K.}~\bibnamefont
  {Takahashi}},\ }\bibfield  {title} {\emph {\bibinfo {title} {Controlling and
  exploring quantum systems by algebraic expression of adiabatic gauge
  potential},\ }}\href {\doibase 10.1103/PhysRevA.103.012220} {\bibfield
  {journal} {\bibinfo  {journal} {Phys. Rev. A}\ }\textbf {\bibinfo {volume}
  {103}},\ \bibinfo {pages} {012220} (\bibinfo {year} {2021})}\BibitemShut
  {NoStop}%
\bibitem [{\citenamefont {Chandarana}\ \emph {et~al.}(2022)\citenamefont
  {Chandarana}, \citenamefont {Hegade}, \citenamefont {Paul}, \citenamefont
  {Albarr\'an-Arriagada}, \citenamefont {Solano}, \citenamefont {del Campo},\
  and\ \citenamefont {Chen}}]{approTQD2}%
  \BibitemOpen
  \bibfield  {author} {\bibinfo {author} {\bibfnamefont {P.}~\bibnamefont
  {Chandarana}}, \bibinfo {author} {\bibfnamefont {N.~N.}\ \bibnamefont
  {Hegade}}, \bibinfo {author} {\bibfnamefont {K.}~\bibnamefont {Paul}},
  \bibinfo {author} {\bibfnamefont {F.}~\bibnamefont {Albarr\'an-Arriagada}},
  \bibinfo {author} {\bibfnamefont {E.}~\bibnamefont {Solano}}, \bibinfo
  {author} {\bibfnamefont {A.}~\bibnamefont {del Campo}}, \ and\ \bibinfo
  {author} {\bibfnamefont {X.}~\bibnamefont {Chen}},\ }\bibfield  {title}
  {\emph {\bibinfo {title} {Digitized-counterdiabatic quantum approximate
  optimization algorithm},\ }}\href {\doibase 10.1103/PhysRevResearch.4.013141}
  {\bibfield  {journal} {\bibinfo  {journal} {Phys. Rev. Research}\ }\textbf
  {\bibinfo {volume} {4}},\ \bibinfo {pages} {013141} (\bibinfo {year}
  {2022})}\BibitemShut {NoStop}%
\end{thebibliography}%

\end{document}